\documentclass[12pt]{article}
\input epsf   

\usepackage{amssymb}
\usepackage{latexsym}

\def \be{\begin{equation}}
\def \ee{\end{equation}}
\def \berr{\begin{eqnarray}}
\def \err{\end{eqnarray}}
\def \nn{\nonumber}
\def \a{\alpha}

\def \b{\beta}
\def \g{\gamma}
\def \d{\delta}

\def \Del{\Delta}

\def \L{\Lambda}

\def \eps{\varepsilon}

\def \om{\omega}
\def \l{\lambda}

\def \hf{\frac 12}

\def \p{\varphi}

\def \tw {{\om}}

\def \A{{\cal A}}

\def \F{{\cal F}}
\def \G{{\cal G}}

\def \H{{\cal H}}
\def \S{{\cal S}^2}

\def \RR{{\cal R}}

\def \Tr{\mbox{Tr}}

\def \({\left(}
\def \){\right)}
\def \<{\langle}
\def \>{\rangle}
\def \[{\left[}
\def \]{\right]}

\def \obar{\overline}

\def\tens{\mathop{\otimes}}

\newcommand{\tr}{\triangleright}
\newcommand{\tl}{\triangleleft}

\newcommand \one{{\bf 1}}

\def\reps{representations }
\def\rep{representation }

\def\mf#1{{\mathbb #1}}

\def\R{{\mf{R}}}
\def\C{{\mf{C}}}
\newcommand \reals{\R}
\newcommand \compl{\C}

\def\smash{\mbox{$\,\rule{0.3pt}{1.1ex}\!\times\,$}}

\newcommand{\sect}[1]{\setcounter{equation}{0}\section{#1}}

\newtheorem{prop}{Proposition}[section]
\newtheorem{theorem}[prop]{Theorem}
\newtheorem{lemma}[prop]{Lemma}

\evensidemargin \oddsidemargin
\begin{document}
\begin{titlepage}
\begin{flushright}
LMU-TPW 00-13 \\
UWThPh-2000-19\\
LPTE Orsay  00/54 \\

\end{flushright}
\begin{center}
{\Large \bf Field Theory on }\\
{\Large \bf the $q$--deformed Fuzzy Sphere I}\\[20pt]
H. Grosse$^a$\footnote{grosse@doppler.thp.univie.ac.at}, 
   J.\  Madore$^b$\footnote{John.Madore@th.u-psud.fr} 
and  
H.\  Steinacker$^c$\footnote{Harold.Steinacker@physik.uni-muenchen.de} \\[2ex] 
{\small\it 
        ${}^a$Institut for Theoretical Physics, University of Vienna,\\
              Boltzmanngasse 5, A-1090 Vienna, Austria \\[2ex]

        ${}^b$Max-Planck-Institut f\"ur Physik\\
           F\"ohringer Ring 6, D-80805 M\"unchen  \\
        and \\
        Laboratoire de Physique Th\'eorique et Hautes Energies\\
        Universit\'e de Paris-Sud, B\^atiment 211, F-91405 Orsay \\[2ex]

       ${}^c$Sektion Physik der Ludwig--Maximilians--Universit\"at M\"unchen\\
        Theresienstr.\ 37, D-80333 M\"unchen  \\[1ex] }

{\bf Abstract} \\
\end{center}

We study the $q$--deformed fuzzy sphere, which is related to $D$-branes 
on $SU(2)$ WZW models, for both real $q$ and $q$ a root 
of unity. We construct for both cases a differential calculus which 
is compatible with the star structure, study the integral, and
find a canonical frame of one--forms.
We then consider actions for scalar field theory, as well as for 
Yang--Mills and Chern--Simons--type gauge theories. The zero curvature 
condition is solved.

\noindent
 
\vspace{3cm}
 \vfill
PACS classification 11.10.Kk, 03.65.Fd \\
MSC 81R50, 81T13, 58B30


\end{titlepage}

\sect{Introduction}

There has been considerable work aimed at formulating
models of quantum field theory on non--commutative spaces. The motivation
is to obtain new insights into the UV--divergences and the problem of
renormalization. On some simple noncommutative spaces, it is now possible
to formulate quantum field theories. In some case, the UV divergences 
are completely regularized \cite{grosse,madore_sphere}, while in others 
they persist \cite{filk,johnandme}.
Moreover, it was realized that such noncommutative spaces are in fact
induced by certain sectors of string theory, particularly open strings
ending on $D$--branes with a background $B$ field 
\cite{chu}. This is 
both a valuable source of physical insights, as well as a vindication 
of a more ``puristic'' approach of studying such spaces per se.
In particular, spaces with  quantum group symmetries have also been 
studied from a more formal approach. While quantum groups appear naturally
in the context of 2--dimensional conformal field theories \cite{sierra},
a formulation of a quantum field theory based on such spaces has 
proved to be difficult. 

Recently, Alekseev, Recknagel and Schomerus \cite{alekseev} 
have found that 
spherical $D$--branes in the $SU(2)$ WZW model are seen by open strings
ending on them (in an appropriate background)
as certain quasi--associative algebras, which are closely related to
$q$--deformations of fuzzy spheres. Here $q$ is related to 
the level $k$ of the WZW model by the formula 
\be
q=\exp(\frac{i\pi }{k+2}).
\ee
We shall take this as sufficient motivation to study in detail 
the $q$--deformed fuzzy spheres,
and to formulate field theories on them.

The algebra found in \cite{alekseev} is (weakly) non--associative, 
and covariant under $SU(2)$. Using a  so--called Drinfeld twist, 
it can be transformed into an associative algebra which we call $\S_{q,N}$.
It is covariant under the ``quantum group'' $U_q(su(2))$, which is the 
quantized universal enveloping algebra of Drinfeld and Jimbo 
\cite{drinfeld,jimbo}.
Here $N$ is an integer related to a particular boundary 
condition on the $D$--brane in $SU(2)$ WZW model.

After reviewing the undeformed fuzzy sphere, we define $\S_{q,N}$ 
in Section 2 for both $q\in \R$ and $|q|=1$.
As an algebra, it is simply
a finite--dimensional matrix algebra, equipped with additional structure 
such as an action of $U_q(su(2))$, a covariant differential calculus,
a star structure, and an integral. For $q \in \R$,
this is precisely the ``discrete'' series of Podle\'s spheres \cite{podles}.
The case $|q|=1$, which is most relevant to string theory,
has apparently not been studied in detail in the literature. In
Section 3, we develop the non--commutative
differential geometry on $\S_{q,N}$, using an approach which is suitable
for both $q \in \R$ and $|q|=1$.  The differential calculus
turns out to be rather elaborate, but quite satisfactory.
We are able to show, in particular, that 
in both cases there exists a 3--dimensional exterior differential calculus
with real structure and a Hodge star, and we develop a frame 
formalism \cite{DubKerMad89a,DimMad96,Mad99c}.
This allows us to write Lagrangians for field theories on $\S_{q,N}$.
In particular, the fact that the tangential space is 3--dimensional
unlike in the classical case turns out to be very interesting physically,
and is related to 
recent results \cite{alekseev2} on Chern--Simons actions on the $D$--branes.

Using these tools, we study in Section 4 actions for scalar fields
and abelian gauge fields on $\S_{q,N}$. The latter case is particularly
interesting, since it turns out that certain actions for gauge 
theories arise in a very natural way in terms of polynomials
of one--forms. In particular, the kinetic terms arise automatically
due to the noncommutativity of the space. Moreover, because the 
calculus is 3--dimensional, the gauge field
consists of a usual (abelian) gauge field plus a (pseudo) scalar
in the classical limit. 
This is similar to a Kaluza--Klein reduction.
One naturally obtains analogs of Yang--Mills and 
Chern--Simons actions, again because the calculus is
3--dimensional. In a certain limit where $q=1$, 
such actions were shown to arise from open strings ending
on $D$--branes in the $SU(2)$ WZW model \cite{alekseev2}.
The gauge theory actions for $q \neq 1$ suggest a new
version of gauge invariance, where the gauge ``group'' is 
a quotient $U_q(su(2))/I$, which can be identified with the space 
of functions on the deformed fuzzy sphere. This is 
discussed in Section \ref{subsec:gauge}.

Finally in Section 5, we give the precise relation of
$\S_{q,N}$ to the quasi--associative
algebra of functions on $D$--branes found in \cite{alekseev},
using a Drinfeld--twist.


In this paper, we shall only consider the
first--quantized situation; the second quantization is postponed
to a forthcoming paper \cite{ours_2}. The latter
turns out to be necessary for implementing the symmetry
$U_q(su(2))$ on the space of fields in a fully satisfactory way.

\sect{The $q$--deformed fuzzy sphere}

\subsection{Review of the undeformed case}

We briefly recall the definition of the ``standard'' fuzzy sphere
\cite{madore_sphere,grosse}.
Much information about the standard unit sphere $S^2$ in ${\bf R^3}$
is encoded in the infinite dimensional algebra 
of polynomials generated by 
$\tilde x = (\tilde x_1, \tilde x_2, \tilde x_3) \in {\bf R^3}$ with the 
defining relations
\be
[\tilde x_i, \tilde x_j] = 0, \ \ \ \sum_{i=1}^3 \tilde x_i^2 = r^2
\ee
The algebra of functions on the  fuzzy sphere is defined 
as the finite algebra 
${\cal S}^2_N$ generated by $\hat{x} = (\hat{x}_1, \hat{x}_2, \hat{x}_3)$, with
relations 
\be
[\hat{x}_i, \hat{x}_j ] = i \lambda_N \varepsilon_{ijk} \hat{x}_k \ ,\ \
\sum_{i=1}^3 \hat{x}_i^2 = r^2 \ .
\label{fuzzy_x}
\ee
The real parameter $\lambda_N > 0$ characterizes the non-commutativity. 


These relations are realized in a suitable finite--dimensional 
irreducible unitary representations of the $SU(2)$ group. 
This is most conveniently done using the Wigner-Jordan realization of the
generators $\hat{x}_i \ , i=1,2,3$, in terms of two pairs of annihilation and
creation operators $A_{\alpha} , {A^+}^{\alpha} , \alpha = \pm \hf$, 
which satisfy
\be
[A_{\a} , A_{\b} ] = [{A^+}^{\a} , {A^+}^{\b} ] = 0 \ ,\ \
[A_{\a} , {A^+}^{\b} ] = \delta_{\a}^{\b} \ ,
\label{A_cr}
\ee
and act on the Fock space $\cal{F}$ spanned by the vectors
\be
|n_1 ,n_2 \rangle =
\frac{1}{\sqrt{n_1 ! n_2 !}} ({A^+}^{\hf})^{n_1} ({A^+}^{-\hf})^{n_2} 
|0 \rangle \ .
\ee
Here $|0 \rangle$ is the vacuum defined by $A_i |0 \rangle = 0$. 
The operators $\hat{x}_i$ take the form
\be
\hat{x}_i = \frac{\lambda_N}{\sqrt{2}} 
           {A^+}^{\a'} \eps_{\a' \a} \sigma_i^{\a \b} A_{\b}. 
\label{JW}
\ee
Here $\eps_{\a \a'}$ is the antisymmetric tensor 
(spinor metric), and $\sigma_i^{\a \b}$ are the  
Clebsch-Gordan coefficients, 
that is rescaled Pauli--matrices.
The number operator is given by 
$\hat N = \sum_{\a} {A^+}^{\a} A_{\a}$. When restricted to
the $(N+1)$-dimensional subspace
\be
{\cal{F}}_N = \{\sum {A^+}^{\a_1} ... \; {A^+}^{\a_N}|0\rangle  \; 
       (N\; \mbox{creation operators})\}.
\ee
it yields for any given $N = 0,1,2,...\ \ $ the irreducible unitary
representation in which the parameters $\lambda_N$ and $r$ are related as
\be
\frac {r}{\lambda_N} = \sqrt{\frac{N}{2} \left( \frac{N}{2} +1\right) } \ .
\ee
The algebra ${\cal S}^2_N$ generated by the $\hat{x}_i$ is clearly the
simple matrix algebra $Mat(N+1)$.
Under the adjoint action of $SU(2)$, it decomposes into the 
direct sum $(1) \oplus (3) \oplus (5) \oplus ... \oplus (2N+1)$
of irreducible \reps of $SO(3)$ \cite{grosse}.

\subsection{The $q$--deformed fuzzy sphere}
\label{subsec:q_fuzzy}

The fuzzy sphere $\S_N$ is invariant under the action of $SO(3)$, 
or equivalently 
under the action of $U(so(3))$. We shall define finite algebras 
$\S_{q,N}$ generated by ${x}_i$ for $i=1,0,-1$,
which have completely analogous properties to
those of $\S_{N}$, but which are
covariant under the quantized universal enveloping algebra
$U_q(su(2))$. This will be done for both $q \in \R$ and $q$ a phase,
including the appropriate reality structure.
In the first case, the $\S_{q,N}$ will turn out to be the  
``discrete series'' 
of Podle\'s' quantum spheres \cite{podles}.
Here we will study them more closely from the above point of view.
However, we also allow $q$ to be a root of unity,
with certain restrictions.
In a twisted form, this case does appear naturally 
on $D$--branes in  the $SU(2)$ WZW model, as was shown in \cite{alekseev}.

In order to make the analogy to the undeformed case obvious,
we shall perform a $q$--deformed Jordan--Wigner 
construction, which is covariant under $U_q(su(2))$. To fix the notation, we 
recall the basic relations of $U_q(su(2))$
\berr                 
\[H, X^{\pm}\] &=& \pm 2 X^{\pm},  \nonumber \\
\[X^+, X^-\]   &=&  \frac{q^{H}-q^{-H}}{q-q^{-1}}
                    =  [H]_{q}
\label{U_q_rel}
\err    
where the $q$--numbers are defined as 
$[n]_q = \frac {q^n-q^{-n}}{q-q^{-1}}$. 
The action of $U_q(su(2))$ on a tensor product of \reps 
is encoded in the coproduct\footnote{We use the opposite 
coproduct than in the standard conventions, but nevertheless the 
invariant tensors and $\hat R$ --matrices will be the standard
ones. The reason for this is explained in Appendix A.}
\berr  
\Del(H)       &=& H \tens 1 + 1 \tens H \nonumber \\
\Del(X^{\pm}) &=&  X^{\pm} \tens q^{-H/2} + q^{H/2}\tens X^{\pm}.
\label{coproduct_X} 
\err 
The antipode and the counit are given by
\berr
S(H)    &=& -H, \quad  
S(X^+)  = -q^{-1} X^+, \quad S(X^-)  = -q X^-, \nonumber \\
\eps(H) &=& \eps(X^{\pm})=0.
\err
The star structure is related to the
Cartan--Weyl involution $\theta(X^{\pm}) = X^{\mp}, \; \theta(H) = H$,
and will be discussed below.
All symbols will now be understood to carry a label ``$q$'',
which we shall omit.

An algebra $\A$ is called an $U_q(su(2))$--module algebra if 
there exists an action 
\berr
U_q(su(2)) \times \A &\rightarrow& \A, \nn\\
  (u,a) &\mapsto& u \tr a
\err
which satisfies $u \tr (ab) = (u_{(1)} \tr a) (u_{(2)} \tr b)$
for $a, b \in \A$. Here $\Delta(u) = u_{(1)} \tens u_{(2)}$
is the Sweedler notation for the coproduct. 

Consider $q$--deformed creation and anihilation operators 
$A_{\a} , {A^+}^{\a}$ for $ \alpha = \pm \hf$, which
satisfy the relations (cp. \cite{WZ}) 
\berr
{A^+}^{\a} A_{\b} &=& \d^{\a}_{\b} + q \hat{R}^{\a \g}_{\b\d}
                               A_{\g}  {A^+}^{\d} \nn\\
(P^-)^{\a \b}_{\g\d} A_{\a} A_{\b} &=& 0 \nn\\
(P^-)^{\a \b}_{\g\d} {A^+}^{\d} {A^+}^{\g} &=& 0
\label{AA_CR} 
\err  
where $\hat{R}^{\a \g}_{\b\d} = q (P^+)^{\a \g}_{\b\d} - q^{-1}  
(P^-)^{\a \g}_{\b\d}$ is the decomposition of the $\hat R$ --matrix
of $U_q(su(2))$ into the projection operators on the symmetric  
and antisymmetric part. They can be written as  
\berr
(P^-)^{\a \b}_{\g\d}  &=& \frac 1{-[2]_q}\eps^{\a \b} \eps_{\g\d}, \nn\\
(P^+)^{\a \b}_{\g\d}  &=& \sigma_i^{\a \b} \sigma^i_{\g \d}
\label{sigma_completeness}
\err
Here $\eps_{\a \b}$ is the $q$--deformed invariant
antisymmetric tensor, and $\sigma^i_{\a \b}$ are the  $q$--deformed
Clebsch--Gordan coefficients; they are given explicitly in the Appendix. 
The factor $-[2]_q^{-1}$ arises from the relation
$\eps^{\a \b}\eps_{\a\b} = -[2]_q$. 
 The above relations are covariant under $U_q(su(2))$,  
and define a left $U_q(su(2))$--module algebra. 
We shall denote the action on the generators with lower indices by  
\be 
u \tr A_{\a} = A_{\b}\; \pi^{\b}_{\a}(u),  
\ee  
so that $\pi^{\a}_{\b}(u v ) = \pi^{\a}_{\g}(u) \pi^{\g}_{\b}(v)$ for  
$u,v \in U_q(su(2))$.
The generators with upper indices transform in the contragredient  
representation, which means that  
\be 
A^+_{\a} := \eps_{\a \b} {A^+}^{\b} 
\ee 
transforms in the same way under $U_q(su(2))$ as $A_{\a}$.

We consider again the corresponding 
Fock space $\cal{F}$ generated by the  ${A^+}^{\a}$ acting on the   
vacuum $|0\rangle$, and its sectors 
\be 
{\cal{F}}_N = \{\sum {A^+}^{\a_1} ... \; {A^+}^{\a_N}|0\rangle  \; 
       (N\; \mbox{creation operators})\}.
\label{F_N}
\ee
It is well--known that these subspaces ${\cal{F}}_N$
are $N+1$--dimensional, as they are when $q=1$, 
and it follows that they form irreducible \reps 
of $U_q(su(2))$ (at root of unity, this will be true due to the restriction 
(\ref{q_phase}) we shall impose). 
This will be indicated by writing ${\cal{F}}_N = (N+1)$, and 
the decomposition of ${\cal F}$ into irreducible representations is
\be
{\cal F} = {\cal{F}}_0 \oplus {\cal{F}}_1 \oplus {\cal{F}}_2 \oplus ... = 
            (1) \oplus (2) \oplus (3) \oplus ... 
\label{F_decomp}
\ee
Now we define 
\be
\hat Z_i = {A^+}^{\a'} \eps_{\a \a'} \sigma_i^{\a \b} A_{\b} 
\label{Z_i}
\ee   
and  
\be
\hat N = \sum_{\a} {A^+}^{\a'}\eps_{\a \a'} \eps^{\a\b}A_{\b}.  
\ee
After some calculations, 
these operators can be shown to satisfy the relations
\berr
&&\eps_k^{i j} \hat Z_i \hat Z_j =
 \frac{q^{-1}}{\sqrt{[2]_q}}(q^{-1} [2]_q - \l\hat N) \hat Z_k 
\label{ZZ_sigma}\\
&&\hat Z^2 := g^{i j} \hat Z_i \hat Z_j =
            q^{-2}\frac{[2]_q + \hat N}{[2]_q} \hat N 
\label{ZZ_square}
\err
Here $\l = (q-q^{-1})$, $g^{ij}$ is the $q$--deformed invariant tensor 
for spin 1 representations, 
and $\eps_k^{ij}$ is the corresponding $q$--deformed Clebsch--Gordan 
coefficient; they are given in the Appendix.
Moreover, one can verify that
\berr
\hat N{A^+}^{\a} &=& q^{-3}{A^+}^{\a} + q^{-2} {A^+}^{\a} \hat N, \nn\\
\hat N{A}_{\a}   &=& -q^{-1}{A}_{\a} + q^{2} {A}_{\a} \hat N,
\err
which implies that
$$
[\hat N , \hat Z_i ] = 0.
$$
On the subspace ${\cal F}_N$, the ``number'' operator  $\hat N$ 
takes the value 
\be
\hat N {\cal F}_N = q^{-N-2} [N]_q {\cal F}_N
\label{hat_N}
\ee
It is convenient to introduce also an undeformed number operator
$\hat n$ which has eigenvalues
$$
\hat n \F_N = N \F_N,
$$
in particular $\hat n A_{\a} = A_{\a} (\hat n-1)$.

On the subspaces ${\cal F}_N$, the relations (\ref{ZZ_sigma}) become
\berr
\eps_k^{i j}  x_i  x_j &=& 
            \L_N \; x_k, \label{xx_sigma}\\ 
 x\cdot x &:=& g^{i j} x_i x_j =  r^2. 
\label{xx_square}
\err
Here the variables have been rescaled to $x_i$ with  
$$
x_i = r \; \frac{q^{\hat n +2}}{\sqrt{[2]_q C_N}} \; \hat Z_i.
$$
The $r$ is a real number, and we have defined
\berr
C_N &=& \frac{[N]_q [N+2]_q}{[2]_q^2}, \nn\\
\L_N &=& r \; \frac{[2]_{q^{N+1}}}{\sqrt{[N]_q [N+2]_q}}.
\err
Using a completeness relation (see Appendix A),
(\ref{xx_sigma}) can equivalently be written as 
\be
(P^-)^{ij}_{kl}  x_i  x_j  = \frac 1{[2]_{q^2}}\; \L_N  \eps^n_{kl} \; x_n.
\ee
There is no $i$ in the commutation relations, because we use
a weight basis instead of Cartesian coordinates.
One can check that these relations precisely reproduce the 
``discrete'' series 
of Podle\'s' quantum spheres 
(after another rescaling), see \cite{podles}, Proposition 4.II.
Hence we define $\S_{q,N}$ to be 
the algebra generated by the variables $x_i$ acting on ${\cal F}_N$.
Equipped with a suitable star structure and a differential structure, 
this will be the $q$--deformed fuzzy sphere.

It is easy to see that the algebra $\S_{q,N}$ 
is simply the full matrix algebra $Mat(N+1)$, i.e.
it is the same algebra as $\S_{N}$ for $q=1$.
This is because ${\cal{F}}_N$ is an irreducible \rep
of $U_q(su(2))$. To see it, we use complete reducibility \cite{rosso}
of the space of polynomials in $x_i$ of degree $\leq k$ to conclude that 
it decomposes into the direct sum of irreducible \reps
$(1) \oplus (3) \oplus (5) \oplus ... \oplus (2k+1)$.
Counting dimensions and noting that $x_1^{N} \neq 0 \in (2N+1)$,
it follows that $\dim(\S_{q,N}) = (N+1)^2 = \dim Mat(N+1)$, and hence
\be
\S_{q,N} = (1) \oplus (3) \oplus (5) \oplus ... \oplus (2N+1).
\label{A_decomp}
\ee
This is true even if $q$ is a root of unity provided 
(\ref{q_phase}) below holds, a relation which will be necessary 
for other reasons as well. This is the decomposition of the functions 
on the $q$--deformed fuzzy sphere into $q$--spherical harmonics, and 
it is automatically truncated.
Note however that 
not all information about a (quantum) space 
is encoded in its algebra of functions; in addition, one must 
specify for example a differential calculus and symmetries.
For example, the action of $U_q(su(2))$ on $\S_{q,N}$ is different
from the action of $U(su(2))$ on $\S_{N}$.

The covariance of $\S_{q,N}$ under $U_q(su(2))$ can also be 
stated in terms of the
quantum adjoint action. It is convenient to consider the  
cross--product algebra $U_q(su(2)) \smash \S_{q,N}$, which 
as a vector space is equal to $U_q(su(2)) \tens \S_{q,N}$,
equipped with an algebra structure defined by
\be
ux = (u_{(1)}\tr x)u_{(2)}.
\label{smash}
\ee
Here the $\tr$  denotes the action of
$u \in U_q(su(2))$ on $x \in \S_{q,N}$.
Conversely, the action of $U_q(su(2))$ on $\S_{q,N}$ 
can be written as $u \tr x = u_{(1)} x Su_{(2)}$.
The relations (\ref{smash}) of $U_q(su(2)) \smash \S_{q,N}$
are automatically realized on the \rep $\F_N$.

Since both algebras $\S_{q,N}$ and $U_q(su(2))$ act on $\F_N$ and 
generate the full matrix algebra 
$Mat(N+1)$, it must be possible to express the generators 
of $U_q(su(2))$ in terms of the $\hat Z_i$. 
The explicit relation can be obtained by 
comparing the relations (\ref{xx_sigma}) with (\ref{smash}). One finds
\berr
X^+ q^{-H/2} &=& q^{N+3} \;\hat Z_1,     \nn\\
X^- q^{-H/2} &=& -q^{N+1}\; \hat Z_{-1}, \nn\\
q^{-H} &=& \frac{[2]_{q^{N+1}}}{[2]_q} + 
              \frac{q^{N+2}(q-q^{-1})}{\sqrt{[2]_q}}\; \hat Z_0,
\label{U_A_relation}
\err
if acting on $\F_N$. In fact, this defines an algebra map
\be
j:\quad U_q(su(2)) \rightarrow \S_{q,N}
\label{j_map}
\ee
which satisfies
\be
j(u_{(1)}) x j(Su_{(2)}) = u \tr x
\label{q_adjoint}
\ee
for $x \in \S_{q,N}$ and $u \in U_q(su(2))$. This is analogous to 
results in \cite{fiore_co,schraml}. We shall often omit $j$ from 
now on. In particular, $\S_{q,N}$ is 
the quotient of the algebra $U_q(su(2))$ by the relation 
(\ref{xx_square})\footnote{more precisely, its finite--dimensional 
representation.}.
The relations (\ref{q_adjoint}) and those of $U_q(su(2))$ can be verified
explicitly using (\ref{ZZ_sigma}). Moreover, one can verify that 
it is represented correctly on $\F_N$
by observing that $X^+ (A^+_{1/2})^N |0\rangle = 0$, which means that
$(A^+_{1/2})^N |0\rangle $ is the highest--weight vector of $\F_N$.

\subsection{Reality structure for $q \in \R$ }

In order to define a real quantum space, we must also construct a star
structure, which is an involutive anti--linear anti--algebra map. 
For real $q$, the algebra
(\ref{AA_CR}) is consistent with the following star structure
\berr
(A_{\a})^{\ast} &=& {A^+}^{\a}  \nn\\
({A^+}^{\a})^{\ast} &=& A_{\a}
\label{A_star_real}
\err
This can be verified using the standard compatibility relations
of the $\hat R$ --matrix with the invariant tensor \cite{FRT}.
On the generators $x_i$, it implies the relation
\be
x_i^{\ast} = g^{ij} x_j,
\label{x_star_real}
\ee
as well as the equality
$$
\hat N^{\ast} = \hat N.
$$
The algebras $\S_{q,N}$ are now precisely 
Podle\'s' ``discrete'' $C^*$ algebras $\tilde{S}^2_{q,c(N+1)}$.
Using (\ref{U_A_relation}), this is equivalent to
\be
H^{\ast} = H, \; (X^{\pm})^{\ast} = X^{\mp},
\label{U_star}
\ee
which is the star--structure for the compact form $U_q(su(2))$.
It is well--known that there is a unique Hilbert space structure on 
the subspaces ${\cal F}_N$ such that they are unitary irreducible \reps of 
$U_q(su(2))$. Then the above star is simply the operator adjoint.

\subsection{Reality structure for $q$ a phase}
\label{subsec:real_form}

When $q$ is a phase, finding the correct star structure is not quite
so easy. The difference with the case $q \in \R$ is that 
$\Delta(u^{\ast}) = (\ast \tens \ast) \Delta'(u)$
for $|q|=1$ and $u \in U_q(su(2))$, 
where $\Delta'$ denotes the flipped coproduct. 
We shall define a star only on the algebra
$\S_{q,N}$ generated by the $x_i$, and not on the full
algebra generated by $A_{\a}$ and ${A_{\a}^+}$.

There appears to be an obvious choice at first sight, namely
$x_i^{*} = x_{i}$, which is indeed 
consistent with (\ref{xx_sigma}). However, it is the
wrong choice for our purpose, because it induces the noncompact 
star structure $U_q(sl(2,\R))$. 

Instead, we define a star--structure on $\S_{q,N}$ as follows. 
The algebra
$U_q(su(2))$ acts on the space $\S_{q,N}$, which generically
decomposes as $(1)\oplus (3) \oplus ...\oplus (2N+1)$. 
This decomposition should be a direct sum of {\em unitary} \reps  
of the compact form 
of $U_q(su(2))$, which means that the star structure 
on $U_q(su(2))$ should be (\ref{U_star}), as it is for real $q$.
There is a slight complication, because
not all finite--dimensional irreducible \reps are 
unitary if $q$ is a phase \cite{keller}.
However, all \reps with dimension $\leq 2N+1$ are
unitary provided $q$ has the form
\be
q = e^{i\pi \p}, \quad \mbox{with} \;\; \p < \frac 1{2N}.
\label{q_phase}
\ee
This will be assumed from now on.

As was pointed out before, we can consider the algebra  $\S_{q,N}$
as a quotient of $U_q(su(2))$ via (\ref{U_A_relation}). It acts on
${\cal F}_N$, which is an irreducible \rep of $U_q(su(2))$, and hence 
has a natural Hilbert space structure.  We define the 
star on the operator algebra $\S_{q,N}$ by the adjoint
(that is by the matrix adjoint in an orthonormal basis), hence by the 
star (\ref{U_star}) using the identification (\ref{U_A_relation}).

There is a very convenient way to write down this star structure
on the generators $x_i$, similar as in \cite{ads_space}. 
It involves an element $\tw$ of an extension of $U_q(su(2))$ introduced by 
\cite{kirill_resh} and \cite{lev_soib},
which implements the Weyl reflection on irreducible representations.
The essential properties are
\berr
\Del(\tw) &=& \RR^{-1}\tw\tens \tw,        \\ 
\tw u \tw^{-1} &=& \theta S^{-1}(u),   \label{S_theta}\\
\tw^2 &=& v \epsilon,
\err
where $v$ and $\epsilon$ are central elements in $\in U_q(su(2))$ 
which take the values $q^{-N(N+2)/2}$ resp. $(-1)^{N}$ on $\F_N$. 
Here $\RR = \RR_1 \tens \RR_2 \in U_q(su(2)) \tens U_q(su(2))$ is the 
universal $R$ element. In a suitable (weight) basis of a
unitary representation of $U_q(su(2))$, the matrix representing $\tw$ 
is given the invariant tensor, 
$\pi^i_j(\tw) = - q^{-N(N+2)/4} g^{ij}$, and $\tw^{\ast} = \tw^{-1}$. 
This is discussed in detail in \cite{ads_space}. From now on, we denote
with $\tw$ the element in $\S_{q,N}$ which represents this element 
on $\F_N$.

We claim that the star structure 
on $\S_{q,N}$ as explained above is given by the following 
formula:
\be
x_i^{\ast} = - \tw x_i \tw^{-1} \
        = x_j {L^-}^{j}_{k} q^{-2} g^{k i},
\label{x_star_phase}
\ee
where
\be
{L^-}^{i}_{j} = \pi^{i}_{j}(\RR_1^{-1}) \RR_2^{-1}           \label{L_-}
\ee
as usual \cite{FRT}; a priori, ${L^-}^{i}_{i} \in U_q(su(2))$, 
but it is understood here as an element of $\S_{q,N}$ via 
(\ref{U_A_relation}).
One can easily verify  using $(\eps^{ij}_k)^* = - \eps^{ji}_k$ (for $|q|=1$)
that (\ref{x_star_phase}) is consistent with the relations 
(\ref{xx_sigma}) and (\ref{xx_square}). In the limit $q \rightarrow 1$, 
${L^-}^{i}_{j} \rightarrow \d^{i}_{j}$, therefore (\ref{x_star_phase})
agrees with (\ref{x_star_real}) in the classical limit.
Hence we define the $q$--deformed fuzzy sphere for $q$ a phase
to be the algebra $\S_{q,N}$ equipped with 
the star--structure (\ref{x_star_phase}).

To show that (\ref{x_star_phase}) is correct in the sense explained above, 
it is enough to 
verify that it induces the star structure (\ref{U_star}) on $U_q(su(2))$, 
since both $U_q(su(2))$ and $\S_{q,N}$ generate 
the same algebra $Mat(N+1)$. This can easily be seen 
using (\ref{S_theta}) and (\ref{U_A_relation}).
A somewhat related conjugation has been proposed in 
\cite{ads_space,mack_schom} using the universal element
$\RR$.

\subsection{Invariant integral}

The integral on $\S_{q,N}$ is defined to be the 
unique functional on $\S_{q,N}$ which is invariant 
under the (quantum adjoint) action of $U_q(su(2))$.
It is given by the projection on the 
trivial sector in the decomposition (\ref{A_decomp}). We claim that it can be 
written explicitly using the quantum trace:
\be
\int\limits_{\S_{q,N}} f(x_i) := 4\pi r^2 \frac 1{[N+1]_q} \Tr_q(f(x_i)) = 
              4\pi r^2  \frac 1{[N+1]_q} \Tr(f(x_i) \;q^{-H})
\label{integral}
\ee
for $f(x_i) \in \S_{q,N}$, where the trace is taken on $\F_N$. 
Using $S^{-2} (u) = q^{H} u q^{-H}$
for $u \in U_q(su(2))$, it follows that 
\be
\int\limits_{\S_{q,N}}f g = \int\limits_{\S_{q,N}} S^{-2}(g) f.
\label{cyclic_integral}
\ee
This means that it is indeed 
invariant under the quantum adjoint action, 
\berr
\int\limits_{\S_{q,N}} u \tr f(x_i) &=& \int\limits_{\S_{q,N}} 
                                            u_1 f(x_i)S(u_2) \nn\\
    &=& \int\limits_{\S_{q,N}} S^{-1}(u_2) u_1 f(x_i)  
    = \eps(u) \; \int\limits_{\S_{q,N}} f(x_i),
\label{invariance_int}
\err
using the identification (\ref{U_A_relation}).
The normalization constant is obtained from
$$
\Tr_q(1) = \Tr(q^{-H}) = q^{N} + q^{N-2} + ... + q^{-N} = [N+1]_q
$$
on $\F_N$, so that $\int\limits_{\S_{q,N}} 1 =4\pi r^2$.

\begin{lemma}
Let $f \in \S_{q,N}$. Then 
\be
\Big(\int\limits_{\S_{q,N}} f\Big)^* = \int\limits_{\S_{q,N}}f^{\ast}
\label{I_star_real}
\ee
for real $q$, and 
\be
\Big(\int\limits_{\S_{q,N}}f\Big)^* = \int\limits_{\S_{q,N}} f^{\ast} q^{2H}
\label{I_star_phase}
\ee
for $q$ a phase, with the appropriate star structure 
(\ref{x_star_real}) respectively (\ref{x_star_phase}). 
In (\ref{I_star_phase}), we use (\ref{U_A_relation}).
\end{lemma}

\begin{proof}
Assume first that $q$ is real, and consider the functional
$$
I_{q,N}(f) :=  \Tr(f^{\ast}q^{-H})^*
$$
for $f \in \S_{q,N}$. Then
\berr
I_{q,N}(u \tr f) &=& \Tr((u_1 f S(u_2))^{\ast}\; q^{-H})^* \nn\\ 
   &=& \Tr (S^{-1}((u^{\ast})_2) f^{\ast}
       (u^{\ast})_1 \; q^{-H})^* 
      = \Tr (f^{\ast}(u^{\ast})_1 S((u^{\ast})_2) q^{-H})^* \nn\\
    &=& \eps(u) \;  I_{q,N} (f),
\err
where $(S(u))^{\ast} = S^{-1}(u^{\ast})$ and 
$(\ast\tens\ast) \Delta(u) = \Delta(u^{\ast})$
was used. Hence $I_{q,N}(f)$ is invariant as well, 
and (\ref{I_star_real}) follows  
using uniqueness of the integral (up to normalization).
For $|q| = 1$, we define 
$$
\tilde I_{q,N}(f) :=  \Tr(f^{\ast}q^H)^*
$$
with the star structure (\ref{x_star_phase}). 
Using $(S(u))^{\ast} = S(u^{\ast})$ and 
$(\ast\tens\ast) \Delta(u) = \Delta'(u^{\ast})$,
an analogous calculation shows that $\tilde I_{q,N}$ is invariant under 
the action of $U_q(su(2))$, which again implies (\ref{I_star_phase}).
\end{proof}

For  $|q|=1$, the integral is neither real nor positive, hence
it cannot be used for a GNS construction. Nevertheless, it is clearly the 
appropriate functional to define an action for field theory, since it is
invariant under $U_q(su(2))$. To find a way out, we 
introduce an auxiliary antilinear algebra--map on $\S_{q,N}$ by
\be
\obar{f} = S^{-1}(f^{\ast})
\label{obar}
\ee
where $S$ is the antipode on $U_q(su(2))$, using (\ref{U_A_relation}).
Note that $S$ preserves the relation (\ref{xx_square}), hence it
is well--defined on $\S_{q,N}$. 
This is not a star structure, since
$$
\obar{\obar{f}} = S^{-2} f
$$
for $|q|=1$. Using (\ref{U_A_relation}), one finds in particular
\be
\obar{x_i} = - g^{ij} x_{j}.
\ee
This is clearly consistent 
with the relations (\ref{xx_sigma}) and (\ref{xx_square}).
We claim that (\ref{I_star_phase}) can now be stated as
\be
\Big(\int\limits_{\S_{q,N}}f\Big)^* = \int\limits_{\S_{q,N}} \obar{f}
\qquad \mbox{for } \; |q|=1.
\label{I_star_bar}
\ee
To see this, observe first that 
\be
\Tr(S(f)) = \Tr(f),
\ee
which follows either from the fact that $\hat I_{q,N}(f):= \Tr(S^{-1}(f) q^H)
 = \Tr(S^{-1}(q^{-H} f))$
is yet another invariant functional, or using 
$\tw f \tw^{-1} = \theta S^{-1}(f)$ together with the observation that the 
matrix \reps of $X^{\pm}$ in a suitable basis are real. 
This implies 
\be
\Tr_q (f^{\ast} q^{2H}) = 
\Tr(f^{\ast} q^{H}) = \Tr(S(q^{-H} S^{-1}(f^{\ast}))) = 
\Tr(q^{-H} S^{-1}(f^{\ast})) = \Tr_q( \obar{f}), 
\ee
and (\ref{I_star_bar}) follows.
Now we can write down a positive inner product on $\S_{q,N}$:
\begin{lemma}
The sesquilinear forms
\be
(f,g):=  \int\limits_{\S_{q,N}} f^{\ast} g \quad \mbox{for}\;\;  q\in \R
\ee
and
\be
(f,g):=  \int\limits_{\S_{q,N}} \obar{f} g, \quad \mbox{for}\;\;  |q|=1
\ee
are hermitian, that is $(f,g)^* = (g,f)$, and satisfy
\be
(f, u\tr g) = (u^{\ast} \tr f, g)
\ee
for both $q\in\R$ and $|q|=1$. They are positive definite provided
(\ref{q_phase}) holds for $|q|=1$, and 
define a Hilbert space structure on $\S_{q,N}$. 
\label{inner_product_lemma}
\end{lemma}
\begin{proof}
For $q\in \R$, we have 
\berr
(f, u \tr g) &=& \int\limits_{\S_{q,N}} f^{\ast}u_1 g Su_2 = 
                   \int\limits_{\S_{q,N}} S^{-1}(u_2) f^{\ast} u_1 g = 
                   \int\limits_{\S_{q,N}} (S((u^{\ast})_2)^{\ast} 
                     f^{\ast} ((u^{\ast})_1)^{\ast} g  \nn\\
               &=&  \int\limits_{\S_{q,N}}\((u^{\ast})_1 f
                    S(u^{\ast})_2\)^{\ast} g =  (u^*\tr f, g),
\err
and hermiticity is immediate. For $|q|=1$, consider
\berr
(f, u \tr g) &=& \int\limits_{\S_{q,N}} \obar{f}u_1 g Su_2 = 
                  \int\limits_{\S_{q,N}} S^{-1}(u_2) S^{-1}(f^{\ast})u_1 g=
                \int\limits_{\S_{q,N}} S^{-1}\( (u^{\ast})_1 f 
                  S(u^{\ast})_2 \)^{\ast} g \nn\\
              &=&  (u^*\tr f, g).
\err 
Hermiticity follows using (\ref{I_star_bar}):
$$
(f, g)^* = \int\limits_{\S_{q,N}} \obar{\obar{f}} \obar{g} = 
                \int\limits_{\S_{q,N}} S^{-2}(f) \obar{g} = 
                \int\limits_{\S_{q,N}} \obar{g} f = (g, f).
$$
Using the assumption (\ref{q_phase}) for $|q|=1$, 
it is not difficult to see that they are also positive--definite.
\end{proof}

\sect{Differential Calculus}

In order to write Lagrangians, it is convenient to use the notion 
of an (exterior) differential calculus \cite{worono,connes}. 
A covariant differential calculus over $\S_{q,N}$ is a 
graded bimodule $\Omega^*_{q,N} = \oplus_n \; \Omega^n_{q,N}$
over $\S_{q,N}$ which is a $U_q(su(2))$--module algebra, 
together with an exterior derivative $d$ which satisfies $d^2=0$ and 
the graded Leibnitz rule. We define the dimension of
a calculus to be the rank of $\Omega^1_{q,N}$
as a free right $\S_{q,N}$--module.

\subsection{First--order differential forms}
 
Differential calculi for the Podle\'s sphere have been studied before
\cite{podles_calc,schmuedgen}. 
It turns out that 2--dimensional calculi 
do not exist for the cases we are interested in;
however there exists a unique 3--dimensional module of 1--forms.
As opposed to the classical case, it contains an 
additional ``radial'' one--form.
This will lead to an additional scalar field, 
which will be discussed later.

By definition, it must be possible to write any term 
$x_i d x_j$ in the form $\sum_k d x_k f_k(x)$. Unfortunately the
structure of the module of 1--forms turn out to be not quadratic, 
rather the $f_k(x)$ are polynomials of order up to 3.
In order to make it more easily tractable and to find suitable
reality structures, we will construct this calculus using 
a different basis.
First, we will define the bimodule of 1--forms $\Omega^1_{q,N}$ 
over $\S_{q,N}$ which is covariant under 
$U_q(su(2))$, such that $\{d x_i\}_i$  is a free right 
$\S_{q,N}$--module basis, together with a map 
$d: \S_{q,N} \rightarrow \Omega^1_{q,N}$
which satisfies the Leibnitz rule. Higher--order differential forms 
will be discussed below. 

Consider a basis of one--forms $\xi_i$ for $i=-1,0,1$ with the 
covariant commutation
relations\footnote{they are not equivalent to 
$u \xi_i = u_{(1)} \tr \xi_i u_{(2)}$.}
\be
x_i \xi_j = \hat{R}^{k l}_{ij} \xi_k x_l,
\label{xxi_braid}
\ee
using the $(3) \tens (3) $  $\hat{R}$--matrix of $U_q(su(2))$.
It has the projector decomposition
\be
\hat{R}^{k l}_{ij} = q^2 (P^+)^{kl}_{ij} - q^{-2} (P^-)^{kl}_{ij} + 
                     q^{-4} (P^0)^{kl}_{ij},
\ee
where $(P^0)^{kl}_{ij} = \frac 1{[3]_q} g^{kl}g_{ij}$, and 
$(P^-)^{kl}_{ij} = \sum_n \frac 1{[2]_{q^2}}\; \eps^{kl}_n \eps^n_{ij}$.
The relations (\ref{xxi_braid}) are consistent with (\ref{xx_square}) and 
(\ref{xx_sigma}), using the braiding  relations \cite{resh_1}
\berr
\hat{R}^{k l}_{ij} \hat{R}^{rs}_{lu} \eps^{ju}_n &=& 
                     \eps^{kr}_t \hat{R}^{ts}_{in}, \label{RRC_braid} \\
\hat{R}^{k l}_{ij} \hat{R}^{rs}_{lu} g^{ju} &=& g^{kr} \d^{s}_{i}
\label{RRg_braid}
\err
and the quantum Yang--Baxter equation
$\hat{R}_{12}\hat{R}_{23} \hat{R}_{12} = \hat{R}_{23}\hat{R}_{12}\hat{R}_{23}$,
in shorthand--notation \cite{FRT}.
We define $\Omega^1_{q,N}$ to be the free right module over $\S_{q,N}$ 
generated by the $\xi_i$. It is clearly a bimodule
over $\S_{q,N}$. To define the exterior derivative, consider
\be
\Theta:= x \cdot \xi = x_i \xi_j g^{ij},
\ee
which is a singlet under $U_q(su(2))$.  
It turns out (see Appendix B) that 
$[\Theta, x_i] \neq 0 \in  \Omega^1_{q,N}$.
Hence 
\be
df := [\Theta, f(x)]
\label{d_0}
\ee
defines a nontrivial derivation $d: \S_{q,N} \rightarrow \Omega^1_{q,N}$, 
which completes the definition of the calculus up to first order.
In particular, it is shown in Appendix B that
\be
dx_i = - \L_N \eps^{nk}_i x_n \xi_k +
               (q-q^{-1})(q x_i \Theta - {r^2} q^{-1} \xi_i).
\label{dx_i}
\ee
Since all terms are linearly independent, this is a 3--dimensional 
first--order differential calculus, and by the uniqueness 
it agrees with the 3--dimensional calculus in \cite{podles_calc,schmuedgen}.
In view of (\ref{dx_i}), it is not surprising that the commutation relations 
between the generators $x_i$ and $d x_i$ are very complicated 
\cite{schmuedgen}; will not write them down here. The meaning of the
$\xi$--forms will become more clear in Section \ref{subsec:frames}. 

Using (\ref{C_g}) and the relation $\xi \cdot x = q^4 x \cdot \xi$, 
one finds that
$$
x \cdot dx = \(-\L_N^2 +([2]_{q^2} - 2) r^2\) \Theta.
$$
On the other hand, this must be equal to 
$x_i \Theta x_j g^{ij} - r^2 \Theta$,
which implies that
$$
x_i \Theta x_j g^{ij} = \a r^2 \Theta
$$
with
\be
\a = [2]_{q^2} -1 - \frac{\L_N^2}{r^2} =  1-\frac 1{C_N}.
\ee
Combining this, it follows that 
\be
dx \cdot x = r^2\; \frac 1{C_N}\; \Theta = -  x\cdot dx.
\ee  

Moreover, using the identity (\ref{CC_id}) one finds
\be
 \eps_i^{jk} x_j dx_k = (\a-q^2) r^2  \eps_i^{jk} x_j \xi_k - \L_N r^2 \xi_i
                     + q^2 \L_N x_i \Theta,
\label{sxdx_id}
\ee
which together with (\ref{dx_i}) yields
\be 
\xi_i = \frac{q^2}{r^2}\; \Theta x_i 
       + \frac{q^2 C_N \L_N}{r^4} \eps_i^{jk} x_j dx_k
       - q^2 (1-q^2) \frac{C_N}{r^2} dx_i.
\label{xi_i}
\ee

\subsection{Higher--order differential forms}

Podle\'s \cite{podles_calc} has constructed an extension 
of the above 3--dimensional calculus including higher--order forms
for a large class of quantum spheres. This class does not include ours,
however, hence
we will give a different construction based on $\xi$--variables,
which will be suitable for $q$ a phase as well.

Consider the algebra
\be
\xi_i \xi_j = -q^2 \hat R^{kl}_{ij} \xi_k \xi_l
\label{xixi}
\ee
which is equivalent to $(P^+)^{ij}_{kl} \xi_i \xi_j = 0$, 
$(P^0)^{ij}_{kl} \xi_i \xi_j = 0$
where $P^+$ and $P^0$ are the projectors on the symmetric
components of $(3) \tens (3)$ as above; hence the product is totally
($q$--) antisymmetric. 
It is not hard to see (and well--known) that the dimension of the 
space of polynomials of order $n$ in the $\xi$ is 
$(3,3,1)$ for $n=(1,2,3)$, and zero for $n>3$, as classically.
We define $\Omega^n_{q,N}$ to be the free right $\S_{q,N}$--module
with the polynomials of order $n$ in $\xi$ as basis; this is covariant
under $U_q(su(2))$. Then $\Omega^n_{q,N}$ is in fact a (covariant) 
$\S_{q,N}$--bimodule, since the commutation relations (\ref{xxi_braid})
between $x$ and $\xi$ are consistent with (\ref{xixi}), which follows from
the quantum  Yang--Baxter equation.  
There remains to construct the exterior derivative.
To find it, we first note that (perhaps surprisingly) $\Theta^2 \neq 0$,
rather
\be
\Theta^2 = -\frac{q^{-2} \L_N}{[2]_{q^2}} \eps^{ijk} x_i \xi_j \xi_k.
\label{theta_2}
\ee
The $\eps^{ijk}$ is defined in (\ref{q_epsilon}).
By a straightforward but lengthy calculation which is sketched 
in Appendix B, one can show that
\be
dx_i dx_j g^{ij} + \frac {r^2}{C_N} \Theta^2 = 0. \nn
\ee
We will show below that an extension of the calculus to higher--order forms
exists; then this can be rewritten as
\be
d \Theta - \Theta^2 = 0.
\label{dtheta}
\ee
The fact that $\Theta^2 \neq 0$ makes the construction of the extension
more complicated, since now $\a^{(n)} \rightarrow [\Theta,\a^{(n)}]_{\pm}$
does not define an exterior derivative. To remedy this,
the following observation is useful: the map
\berr
\ast_H: \; \Omega^1_{q,N} &\rightarrow& \Omega^2_{q,N}, \nn\\
          \xi_i &\mapsto&  -\frac{q^{-2}\L_N}{[2]_{q^2}}\; \eps_i^{jk} \xi_j \xi_k
\err
defines a left--and right $\S_{q,N}$--module map; in other words, the 
commutation relations between $\xi_i$ and $x_j$ are the same as between 
$\ast_H(\xi_i)$ and $x_j$. This follows from the 
braiding relation (\ref{RRC_braid}). This is in fact the natural analogue
of the Hodge--star on 1--forms in our context, 
and will be discussed further below. Here we note the important identity
\be
\a (\ast_H \b) = (\ast_H \a) \b
\label{star_adj}
\ee
for any $\a, \b \in \Omega^1_{q,N}$, which is proved in Appendix B.
Now (\ref{theta_2}) can be stated as
\be
\ast_H(\Theta) =  \Theta^2,
\label{ast_theta}
\ee
and applying $\ast_H$ to $df = [\Theta, f(Y)]$ one obtains
\be
[\Theta^2, f(x)] =  \ast_H df(x).
\label{comm_theta2}
\ee
Now we define the map 
\berr
d: \; \Omega^1_{q,N} &\rightarrow& \Omega^2_{q,N}, \nn\\
                   \a &\mapsto& [\Theta,\a]_+ - \ast_H(\a).
\label{d_1}
\err
It is easy to see that this defines a graded derivation 
from $\Omega^1_{q,N}$ to $\Omega^2_{q,N}$, and the
previous equation implies immediately that
$$
(d \circ d) f = 0.
$$
In particular,
\be 
d \xi_i = 
(1-q^2) \xi \Theta + \frac{q^{-2} \L_N}{[2]_{q^2}}\eps_i^{jk} \xi_j\xi_k.
\ee
To complete the differential calculus, we extend it to
$\Omega^3_{q,N}$ by
\berr
d: \; \Omega^2_{q,N} &\rightarrow& \Omega^3_{q,N}, \nn\\
                   \a^{(2)} &\mapsto& [\Theta,\a^{(2)}].
\label{d_2}
\err
As is shown in Appendix B, this satisfies indeed
$$
(d \circ d) \a = 0 \quad \mbox{for any} \;\; \a \in \Omega^1_{q,N}.
$$
It is easy to see that the map (\ref{d_2}) is non--trivial.
Moreover 
there is precisely one monomial of order 3 in the $\xi$ variables, given by
\be
\Theta^3 = -\frac{q^{-6}\L_N r^2}{[2]_{q^2}[3]_q}\eps^{ijk} \xi_i \xi_j \xi_k,
\label{om_3}
\ee
which commutes with all functions on the sphere,
\be
[\Theta^3, f] = 0
\label{theta3_f}
\ee
for all $f  \in \S_{q,N}$.
Finally, we complete the definition of the Hodge star operator by 
\be
\ast_H(1) = \Theta^3,
\ee
and by requiring that $(\ast_H)^2 = id$.

\subsection{Star structure}

A $*$--calculus (or a real form of $\Omega^{*}_{q,N}$) is
a differential calculus which is a graded $*$--algebra such that the star 
preserves the grade, and satisfies \cite{worono}
\berr
(\a^{(n)} \a^{(m)})^{\ast} &=& 
          (-1)^{nm} (\a^{(m)})^{\ast} (\a^{(n)})^{\ast}, \nn\\
(d \a^{(n)})^{\ast} &=& d(\a^{(n)})^{\ast}
\label{real_forms}
\err
for $\a^{(n)} \in \Omega^n_{q,N}$; moreover, the action of 
$U_q(su(2))$ must be compatible with the star on $U_q(su(2))$. 
Again, we have to distinguish the cases $q \in \R$ and $|q|=1$.


\paragraph{ \bf 1) $q \in \R$.}
In this case, the star structure must satisfy
\be
(dx_i)^{\ast} =  g^{ij} dx_j, \quad x_i^{\ast} = g^{ij} x_j,
\label{x_calc_real}
\ee
which by (\ref{xx_square}) implies 
\be
\Theta^{\ast} = -\Theta.
\label{theta_real}
\ee
Using (\ref{xi_i}), it follows that
\berr
\!\!\!\xi_i^{\ast} &=& -g^{ij}\xi_j +  
         q^2 (q-q^{-1})\frac{[2]_q C_N}{r^2 } g^{ij} d x_j \nn \\
         &=& -g^{ij}\xi_j  - q^2 (q-q^{-1})\frac{[2]_q C_N}{r^2 }
                   g^{ij}\( \L_N \eps_j^{kl} x_k \xi_l
              - (q-q^{-1})(q x_j \Theta - q^{-1} r^2 \xi_j )\). 
\label{xi_real}
\err
To show that this is indeed compatible with (\ref{xxi_braid}),
one needs the following identity
\be
q^2(q -q^{-1})\frac{[2]_q C_N}{r^2} 
     (dx_i x_j - {\hat{R}}^{k l}_{ij} x_k dx_l)
 = (\one - (\hat{R}^2)^{k l}_{ij}) \xi_k x_l
\label{reality_id}
\ee
which can be verified with some effort, see Appendix B.
In particular, this shows that if one imposed
$x_i \xi_j = (\hat{R}^{-1})^{k l}_{ij} \xi_k x_l$ instead of 
(\ref{xxi_braid}), one would obtain an equivalent calculus.
This is unlike in the flat case, where one has two inequivalent
calculi \cite{fiore,ogievetsky}.
Moreover, one can show that this real form is consistent
with (\ref{xixi}).

\paragraph{\bf 2) $|q| =1$.}
In view of (\ref{x_star_phase}), it is easy to see that  
the star structure in this case is
\be
(\xi_i)^{\ast} =  q^{-4} \tw \xi_i \tw^{-1}, 
               \quad x_i^{\ast} = - \tw x_i \tw^{-1}.
\label{x_calc_phase}
\ee
Recall that $\tw$ is a particular unitary element of $\S_{q,N}$ introduced in 
Section \ref{subsec:real_form}. 

It is obvious using 
$({\hat R}^{k l}_{ij})^* = (\hat{R}^{-1})^{lk}_{ji}$
that this is an involution which is consistent with (\ref{xxi_braid}),
and one can verify that 
\be
\Theta^{\ast} = - \Theta.
\ee
This also implies 
$$
[\tw, \Theta] =0,
$$
hence 
\be
(d x_i)^{\ast} = - \tw d x_i \tw^{-1}.
\ee
Finally, $\ast_H$ is also compatible with the star structure:
\be
(\ast_H(\a))^{\ast} = \ast_H(\a^{\ast})
\ee
where $\a \in \Omega^1_{q,N}$,
for both $q \in \reals$ and $|q| =1$. This is easy to see
for $\a = \xi_i$ in the latter
case, and for $\a = dx_i$ in the case $q \in \reals$.
This implies that indeed
$(d\a^{(n)})^{\ast} = d(\a^{(n)})^{\ast}$
for all $n$.

We summarize the above results:
\begin{theorem}
The definitions (\ref{d_1}), (\ref{d_2}) define a 
covariant differential calculus 
on $\Omega^{*}_{q,N} = \oplus_{n=0}^3 \; \Omega^n_{q,N}$
over $\S_{q,N}$ with $dim(\Omega^n_{q,N}) = (1,3,3,1)$ for $n=(0,1,2,3)$. 
Moreover, this is a $*$--calculus with the 
star structures (\ref{x_calc_real}) 
and (\ref{x_calc_phase}) for $q \in \R$ and $|q|=1$, respectively.
\end{theorem}

\subsection{Frame formalism}
\label{subsec:frames}

On many noncommutative spaces \cite{fiore_co,Mad99c}, it is possible 
to find a particularly 
convenient set of one--forms (a ``frame'') $\theta_a \in \Omega^1$,
which commute with all elements in the function space $\Omega^0$.
Such a frame exists here as well, and in terms of the
$\xi_i$ variables, it takes a similar form to that of \cite{fiore_co}. 
Consider the elements
\berr
\theta^a &=&  \L_N \; S({L^+}^a_j)\; g^{jk} \xi_k \;\; \in \Omega^1_{q,N}, \\
\l_a     &=& \frac 1{\L_N}\; 
             x_i \; {L^+}^i_a \qquad \;\; \in \S_{q,N}.
\label{vielbein}
\err
where as usual
\berr
{L^+}^i_j &=& \RR_1 \pi^i_j(\RR_2),    \label{L_plus} \\
S({L^+}^i_j) &=& \RR_1^{-1} \pi^i_j(\RR^{-1}_2)      
\err
are elements of $U_q(su(2))$, which we
consider here as elements in $\S_{q,N}$ via (\ref{j_map}).
Then the following holds:
\begin{lemma}
\berr
[\theta^a, f] &=& 0,      \label{frame}  \\
d f              &=& [\l_a, f] \theta^a,      \\
\Theta = x_i \xi_j g^{ij} &=& \l_a \theta^a.   \label{Theta_decomp}
\err
for any $f \in \S_{q,N}$. In this sense, 
the $\l_a$ are dual to the frame $\theta^b$. They satisfy the relations
\berr
\l_a \l_b g^{ba} &=& \frac 1{q^{4} \L_N^{2}} \; r^2, \nn\\
\l_a \l_b \eps_c^{ba} &=& -\frac 1{q^{2}} \; \l_c, \nn\\
\theta^a \theta^b  &=& -q^2 \;\hat R^{ba}_{cd}\theta^d \theta^c  \label{theta_AS}\\
d \theta^a &=& \l_b [\theta^a,\theta^b]_+ 
        + \frac 1{q^{2} [2]_{q^2}} \eps^a_{bc} \theta^c \theta^b \nn\\
\ast_H \theta^a    &=& - \frac 1{q^{2}[2]_{q^2}} 
                       \eps^a_{bc} \theta^c \theta^b \nn\\
\theta^a \theta^b \theta^c &=& -\L_N^2 \frac{q^6}{r^2} \;\eps^{cba} \Theta^3
\label{l_relations}
\err
\end{lemma}
In particular in the limit $q =1$, this becomes  
$\l_a = \frac 1{\L_N}\; x_a$,
and $dx_a = - \eps_{ab}^c x_c \theta^b$, using (\ref{dx_i}).

\begin{proof}
Using 
$$
S({L^+}^i_j) x_k = x_l ({\hat R}^{-1})^{ln}_{jk} S({L^+}^i_n)
$$
(which follows from (\ref{smash})) and $\Delta(S({L^+}^i_j)) = 
S({L^+}^n_j) \tens S({L^+}^i_n)$, 
it is easy to check that 
$[\theta^a, x_i] =0$ for all $i,a$, and (\ref{frame}) follows.
(\ref{Theta_decomp}) follows immediately from 
${L^+}^i_a S({L^+}^a_j) = \d^i_j$, and 
To see (\ref{l_relations}), one needs 
the well--known relation ${L^+}^{l}_r {L^+}^k_s g^{sr} = g^{kl}$, as
well as ${L^+}^{l}_r {L^+}^k_s  \eps_n^{sr} = \eps^{kl}_m {L^+}^m_n$;
the latter follows from the quasitriangularity of $U_q(su(2))$. 
The commutation relations among the $\theta$ are obtained 
as in \cite{fiore_co} by observing
\berr
\theta^a \theta^b  &=& 
         \L_N \theta^a  S({L^+}^b_n) g^{nl} \xi_l \nn\\
 &=&     \L_N S({L^+}^b_n) \theta^a g^{nl} \xi_l    \nn\\
 &=&     \L_N^2 S({L^+}^b_n) S({L^+}^a_j) g^{jk}  g^{nl}  \xi_k \xi_l,
\err
using the commutation relations
$\hat R^{kl}_{ij} S{L^+}^i_n  S{L^+}^j_m = 
S{L^+}^k_i  S{L^+}^l_j \hat R^{ij}_{nm}$, as well as (\ref{RRg_braid}).
The remaining relations can be checked similarly.
\end{proof}

\subsection{Integration of forms}
\label{subsec:integral_forms}

As classically, it is natural to define the integral over the forms of
the highest degree, which is $3$ here. Since any $\a^{(3)} \in \Omega^3_{q,N}$
can be written in the from $\a^{(3)} = f \Theta^3$, we define
\be
\int \a^{(3)} = 
\int f \Theta^3 := \int\limits_{\S_{q,N}} f
\ee
by (\ref{integral}), so that $\Theta^3$ is the volume form.
This definition is natural, since $[\Theta^3, f] = 0$.
Integrals of forms with degree $\neq 3$ will be set to zero. 

This integral satisfies an important cyclic property, as did
the quantum trace (\ref{cyclic_integral}).
To formulate it, we extend the map $S^2$ from $\S_{q,N}$ to $\Omega^*_{q,N}$
by 
$$
S^2(\xi_i) = q^{-H} \tr \xi_i,
$$ 
extended as an algebra map. 
Then the following holds (see Appendix B): 
\be
\int \a \; \b = \int S^{-2}(\b)\; \a
\label{cyclic_forms}
\ee
for any $\a, \b \in \Omega^*_{q,N}$ with $deg(\a) + \deg(\b) = 3$.
Now Stokes theorem follows immediately:
\be
\int d\a^{(2)} =  \int [\Theta,\a^{(2)}] = 0
\ee
for any $\a^{(2)} \in  \Omega^2_{q,N}$, because $S^2 \Theta = \Theta$.
This purely algebraic derivation is also valid on some other spaces 
\cite{integral_paper}.

Finally we establish the compatibility of the integral with the star structure.
From $\Theta^{\ast} = -\Theta$ and (\ref{I_star_real}), 
we obtain 
\be
(\int \a^{(3)})^* = -\int (\a^{(3)})^{\ast} \qquad \mbox{for} \;\; q\in \R.
\ee 
For $|q|=1$, we have to extend the algebra map $\obar{f}$ (\ref{obar}) to 
$\Omega^*_{q,N}$. It turns out that the correct definition is
\be
\obar{\xi_i} = - q^{-4}\;g^{ij}\xi_j +  
            q^{-2} (q-q^{-1})\frac{[2]_q C_N}{r^2} g^{ij} d x_j,
\label{obar_xi}
\ee
extended as an antilinear algebra map; 
compare (\ref{xi_real}) for $q\in\R$. To verify that 
this is compatible with (\ref{xxi_braid}) and (\ref{xixi})
requires the same calculations as to verify the star structure 
(\ref{xi_real}) for $q \in \R$. Moreover one can check using 
(\ref{sxdx_id}) that 
\be
\obar{dx_i} = -g^{ij} dx_j,
\ee
which implies that $\obar{\Theta} = \Theta$, and
\berr
\obar{\ast_H(\a)} &=& \ast_H(\obar{\a}), \nn\\
\obar{d\a} &=& d\obar{\a},  \nn\\
\obar{\obar{\a}} &=& S^{-2} \a
\err
for any $\a \in \Omega^*_{q,N}$. Hence we have
\be
(\int \a^{(3)})^* = \int \obar{\a^{(3)}} \qquad \mbox{for} \;\; |q|=1.
\ee

\sect{Actions and fields}

\subsection{Scalar fields}
\label{subsec:scalars}

With the tools provided in the previous sections, it is possible to 
construct actions for 2--dimensional euclidean field theories 
on the $q$--deformed fuzzy sphere. 

We start with scalar fields, which are simply elements 
$\psi \in \S_{q,N}$. 
The obvious choice for the kinetic term is 
\berr
S_{kin}[\psi] &=& i \frac {r^2}{\L_N^2}\int (d\psi)^{\ast} \ast_H d \psi 
                             \qquad \mbox{for} \;\; q\in \R, \nn\\
S_{kin}[\psi] &=&  \frac {r^2}{\L_N^2}\int \obar{d\psi} \ast_H d \psi 
                             \qquad \mbox{for} \;\; |q|=1,
\err
which, using Stokes theorem, can equivalently be written in the form
\berr
S_{kin}[\psi] &=& - i \frac {r^2}{\L_N^2}\int \psi^{\ast} (d \ast_H d) \psi 
  = - \frac {r^2}{\L_N^2} i\int\limits_{\S_{q,N}} \psi^{\ast} (\ast_H d \ast_H d) \psi 
                            \qquad \mbox{for} \;\; q\in \R, \nn\\
S_{kin}[\psi] &=& -  \frac {r^2}{\L_N^2}\int \obar{\psi} (d \ast_H d) \psi 
   =- \frac {r^2}{\L_N^2}\int\limits_{\S_{q,N}} \obar{\psi} (\ast_H d \ast_H d) \psi 
                            \qquad \mbox{for} \;\;  |q|=1. \nn\\
\err
They are real
\be
S_{kin}[\psi]^* = S_{kin}[\psi] 
\ee
for both $q \in \R$ and $|q| =1$, using the reality properties
established in the previous sections. 

The fields can be expanded in terms of the irreducible \reps
\be
\psi(x) = \sum_{K,n} a^{K,n} \; \psi_{K,n}(x) 
\label{psi_field}
\ee
according to (\ref{A_decomp}), with coefficients $a^{K,n} \in \C$;
this corresponds to the first--quantized case. 
However, in order to ensure invariance of the actions
under $U_q(su(2))$ (or a suitable subset thereof),
we must assume that $U_q(su(2))$ acts on products of fields
via the $q$--deformed coproduct. This can be implemented 
consistently only after a ``second quantization'', such that the coefficients
in (\ref{psi_field}) generate a $U_q(su(2))$--module algebra. 
This will be presented in a forthcoming paper \cite{ours_2}.

One can also consider
real fields, which have the form 
\berr
\psi(x)^* &=& \psi(x)  \qquad \mbox{for} \;\; q\in \R, \nn\\
\obar{\psi(x)} &=& q^{H/2}\;\psi(x)\;q^{-H/2}  \qquad \mbox{for} \;\; |q|=1.
\err
This is preserved under the action of a certain real sector 
$\G \subset U_q(su(2))$ (\ref{gauge_group}); 
the discussion is completely parallel to the one 
below (\ref{B_gauge}) in the next section, hence we will not give it here.

Clearly $\ast_H d \ast_H d$ is the analog of the Laplace operator for 
functions, which can also be written in the usual form
$d \d + \d d$, with $\d = \ast_H  d \ast_H$.
It is hermitian by construction. 
We wish to evaluate it on the irreducible \reps 
$\psi_K \in (2K+1)$, that is, on spin--$K$ representations. The result is the 
following:
\begin{lemma}
If $\psi_K \in \S_{q,N}$ is a spin $K$ representation, then 
\be
\ast_H d\ast_H d \psi_{K} = 
   \frac{2}{[2]_q C_N} [K]_q [K+1]_q \; \psi_{K}.
\label{scalar_laplace}
\ee
\label{laplace_lemma}
\end{lemma}
The proof is in Appendix B.

It is useful to write down explicitly the hermitian forms associated to
the above kinetic action. Consider
\berr
S_{kin}[\psi,\psi'] &=& i \frac {r^2}{\L_N^2}\int (d\psi)^{\ast} 
                              \ast_H d \psi'
                             \qquad \mbox{for} \;\; q\in \R, \nn\\
S_{kin}[\psi,\psi'] &=&  \frac {r^2}{\L_N^2}\int \obar{d\psi} \ast_H d \psi'
                             \qquad \mbox{for} \;\; |q|=1.
\err
Using Lemma \ref{inner_product_lemma}, it follows immediately 
that they satisfy 
\berr
S_{kin}[\psi,\psi']^* &=& S_{kin}[\psi',\psi],  \nn\\
S_{kin}[\psi, u\tr \psi']  &=& S_{kin}[u^{\ast} \tr \psi, \psi']
\err
for both $q\in\R$ and $|q|=1$. 
To be explicit, let $\psi_{K,n}$ be an orthonormal basis of (2K+1). We
can be assume that it is a weight basis, so that $n$ labels the weights
from $-K$ to $K$. Then it follows that
\be
S_{kin}[\psi_{K,n},\psi_{K',m}] = c_{K}\; \d_{K,K'}  \;\d_{n,m}
\ee
for some $c_K \in \R$.
Clearly one can also consider interaction terms, which could be of the form
\be 
S_{int}[\psi] = \int\limits_{\S_{q,N}} \psi \psi \psi,
\ee
or similarly with higher degree.

\subsection{Gauge fields}
\label{subsec:gauge}

Gauge theories arise in a very natural way on $\S_{q,N}$. For simplicity,
we consider only the analog of the abelian gauge fields here. 
They are simply one--forms
\be
B = \sum B_a \theta^a r  \quad \in \Omega^1_{q,N},
\label{B_expand}
\ee 
which we expand in terms of the frames $\theta^a$ introduced in Section
\ref{subsec:frames}. Notice that they have 3 independent components, which 
reflects the fact that calculus is 3--dimensional. Loosely speaking, the 
fuzzy sphere does see a shadow of the 3--dimensional embedding space. 
One of the components is essentially radial and should be considered
as a scalar field, however it is naturally tied up with the 
other 2 components of $B$. 
We will impose the reality condition
\berr
B^* &=& B  \qquad \mbox{for} \;\; q\in \R, \nn\\
\obar{B}  &=& q^{H/2}\; B\;q^{-H/2}  \qquad \mbox{for} \;\; |q|=1.
\label{B_real}
\err
Since only 3--forms can be integrated, the most simple candidates for 
Langrangians that can be written down have the form
\be
S_3 = \frac {1}{r^2\L_N^2} \int B^3, \quad 
S_2 = \frac {1}{r^2\L_N^2} \int B \ast_H B, 
    \quad S_4 = \frac {1}{r^2\L_N^2} \int B^2 \ast_H B^2.
\label{B_actions}
\ee
They are clearly real, with the reality condition (\ref{B_real});
the factor $i$ for real $q$ is omitted here. We also define 
\be
F := B^2 - \ast_H B,
\label{curvature}
\ee
for reasons which will become clear below.
The meaning of the field $B$ becomes obvious if one writes it in the form 
\be
B = \Theta + A, \qquad B_a = \frac 1{r}\l_a + A_a
\label{B_A_split}
\ee
While $B$ and $\Theta$ become singular in the limit $N \rightarrow \infty$, 
$A$ remains well--defined. Using 
\berr
F &=& dA + A^2, \nn\\
\int A \Theta^2 &=& \int d A \Theta = \int  \ast_H A \; \Theta,  \nn\\
\int A^2 \Theta &=& \frac 12 \int(A d A + A  \ast_H A)
\err
which follow from (\ref{d_1}), one finds
\berr
S_2 &=& \frac {1}{r^2\L_N^2} \int  A  \ast_H A + 2 A \Theta^2  \nn\\
S_3 &=& \frac {1}{r^2\L_N^2} \int  A^3 + \frac 32(A dA +  A  \ast_H A)  
        + 3 A \Theta^2 +  \Theta^3 
\err
and 
\be 
S_{YM} := \frac {1}{r^2\L_N^2} \int F \ast_H F = 
        \frac {1}{r^2\L_N^2} \int (dA + A^2) \ast_H (dA + A^2).
\label{YM}
\ee
The latter action (which is a linear combination of $S_2, S_3$, and $S_4$) 
is clearly the analog of the Yang--Mills action, 
which in the classical limit contains a gauge field and a scalar,
as we will see below.
In the limit $q \rightarrow 1$, it reduces to the action considered in 
\cite{grosse_madore_schw}.  

The actions $S_3$ and $S_2$ alone contain terms which are linear in $A$, 
which would indicate
that the definition of $A$ (\ref{B_A_split}) is not appropriate. 
However, the linear terms cancel in the following linear combination
\be
S_{CS} := \frac 13 S_3 - \frac 12 S_2 = - \frac{2\pi}{3\L_N^2}
    + \frac 12 \frac {1}{r^2\L_N^2} \int A dA + \frac 23 A^3.
\label{S_CS}
\ee
Notice that the ``mass term'' $A \ast_H A$ has also disappeared. 
This form is clearly the 
analog of the Chern--Simons action. It is very remarkable that it
exists on $\S_{q,N}$, which is related to the fact that the calculus 
is 3--dimensional. In the case $q=1$, 
this is precisely what has been found recently in the context of
2--branes on the $SU(2)$ WZW model \cite{alekseev2}.

In terms of the components (\ref{B_expand}), 
$B^2 = B_a B_b \theta^a \theta^b r^2$, and
$\ast_H B = -\frac {r}{q^2 [2]_{q^2}} B_a \eps^a_{bc}\theta^c\theta^b$. 
Moreover, it is easy to check that 
\berr
\ast_H(\theta^b \theta^c) &=& -q^2 \; \eps^{cb}_a \theta^a,\nn\\
\theta^a \ast_H \theta^b &=& \L_N^2 \frac{q^4}{r^2} g^{ba} \;\Theta^3, \nn\\
\theta^a \theta^b \ast_H \theta^c\theta^d &=& 
      [2]_{q^2}\L_N^2 \frac {q^8}{r^2} 
       (P^-)^{dc}_{a' b'}\; g^{b' b} g^{a' a}\; \Theta^3
   =  \L_N^2 \frac {q^8}{r^2} \eps^{dc}_n \eps^{ba}_m g^{nm} \Theta^3.
\err
Hence 
\berr
F &=&(B_a B_b + \frac 1{q^2 r [2]_{q^2}} B_c\; \eps^c_{ba}) 
          \theta^a \theta^b r^2=
     (\frac{\l_a}{r} A_b + A_a \frac{\l_b}{r} + A_a A_b 
           + \frac 1{q^2 r [2]_{q^2}} A_c \;\eps^c_{ba}) 
           \theta^a \theta^b r^2\nn\\
 &=& F_{ab} \; \theta^a \theta^b r^2,
\label{F_BB}
\err
where we define  $F_{ab}$ to be totally antisymmetric, i.e. 
$F_{ab} = (P^-)^{b' a'}_{ba} F_{a' b'}$ using (\ref{theta_AS}). 
This yields
\berr
S_{YM} &=& q^8 [2]_{q^2}\int\limits_{\S_{q,N}} F_{ab} F_{cd}\; 
                                (P^-)^{dc}_{a' b'}\; g^{b' b} g^{a' a}, \nn\\
\label{gauge_actions_components}
\err

To understand these actions better, we write the gauge fields in terms 
of ``radial'' and ``tangential'' components,
\be
A_a = \frac {x_a}{r}\phi + A^t_a 
\ee
where $\phi$ is defined such that 
\be
x_a A^t_b \; g^{ab} = 0; 
\label{A^t}
\ee
this is always possible. However to get a better insight,
we consider the case $q=1$, and take
the classical limit $N \rightarrow \infty$ in the 
following sense: for a given (smooth) field configuration in $\S_N$, we 
use the sequence of embeddings of $\S_{q,N}$ 
to approximate it for $N \rightarrow \infty$. 
Then terms of the form $[A^t_a, A^t_b]$
vanish in the limit (since the fields are smooth in the limit).
The curvature then splits into a tangential and radial part, 
$F_{ab} = F^t_{ab} + F^{\phi}_{ab}$, where\footnote{the pull--back of
$F$ to the 2--sphere in the classical case is unaffected by this split} 
\berr
F^t_{ab} &=& \frac 1{2r} \([\l_a, A^t_b] - [\l_b, A^t_a]
             + A^t_c \;\eps^c_{ba}\) , \nn\\
F^{\phi}_{ab} 
      &=& \frac 1{2r^2} \(\eps_{ab}^c x_c \phi + [\l_a,\phi] x_b
          - [\l_b, \phi] x_a \).
\err
Moreover,  
\berr
x^a F^t_{ab} &\rightarrow&  \frac 1{4r} [x^a \l_a, A^t_b] 
          - \frac 1{2r} [\l_b, x^a A^t_a] = 0 ,\nn\\
x^a [\l_a,\phi] &\rightarrow& \frac 12[x^a \l_a,\phi] = 0
\label{approx}
\err
in the classical limit, which implies that 
\berr
\int\limits_{S^2} F^t_{ab} {F^{\phi}}^{ab} &=& \int\limits_{S^2} 
          \frac1{2r^2} \eps_{ab}^n x_n \phi {F^t}^{ab}, \nn\\
\int\limits_{S^2} F^{\phi}_{ab} {F^{\phi}}^{ab}
       &=& \int\limits_{S^2} \frac 1{2r^2} 
         \( \phi^2 + [\l_a, \phi] [\l^a, \phi] \) \nn
\err
in the limit. Therefore we find
\be
S_{YM} =  -\int\limits_{S^2} \( 2F^t_{ab} {F^t}_{ab}
         +  \frac2{r^2} \eps_{ab}^n x_n \phi {F^t}^{ab}
         + \frac 1{r^2} (\phi^2 + [\l_a, \phi] [\l^a, \phi] ) \)
\ee
in the limit, as in \cite{grosse_madore_schw}.
Similarly, the Chern--Simons action (\ref{S_CS} ) becomes
\berr
S_{CS}  &\rightarrow& - \frac{2\pi}{3\L_N^2} + 
         \frac {1}{2r^2\L_N^2} \int dA^t (A^t + 2\L_N \Theta\phi)
                 - \L_N^2 \phi^2 \Theta^3   \nn\\
  &=& - \frac{2\pi}{3\L_N^2} + 
      \frac {1}{2r} \int\limits_{S^2} F^t_{ab} (A_c^t + 2\frac{x_c}r \phi)
      \eps^{abc} - \frac {1}{2r^2} \int\limits_{S^2} \phi^2 
\err
for $N \rightarrow \infty$. In the flat limit $r \rightarrow \infty$, the  
term $ F^t_{ab} A_c^t\eps^{abc}$ vanishes because of (\ref{approx}), 
leaving the $F-\phi$ coupling term (after multiplying with $r$).

Back to finite $N$ and $q\neq 1$.
To further justify the above definition of curvature (\ref{curvature}),
we consider the zero curvature condition, $F=0$. 
In terms of the $B$ fields, this is 
equivalent to 
\be
\eps^{ba}_c B_a B_b + \frac 1{q^2 r} B_c = 0
\ee
which is (up to rescaling) the same as equation (\ref{xx_sigma}) with 
{\em opposite} multiplication\footnote{this can be implemented e.g. using the 
antipode of $U_q(su(2))$}; in particular, the solutions 
$B_a \in \S_{q,N}$ are precisely all possible representations of 
$U_q^{op}(su(2))$
in the space of $N+1$--dimensional matrices. They are of course
classified by the number of partitions of $Mat(N+1)$  into blocks with
sizes $n_1, ..., n_k$ such that $\sum n_i = N+1$, as in the case $q=1$.

\paragraph{Gauge invariance.}

We have seen that actions which  describe gauge theories in the limit $q=1$
arise very naturally on $\S_{q,N}$ (as on certain other 
higher--dimensional $q$--deformed spaces \cite{thesis}).
However, it is less obvious in which sense they are actually 
gauge--invariant for $q \neq 1$. 
For $q=1$, the appropriate gauge transformation is
$B \rightarrow U B U^{-1}$, for a unitary element $U \in \S_{N}$. 
This transformation does not work for $q \neq 1$, because of 
(\ref{cyclic_forms}). Instead, we propose the following: let
\berr
\H &=& \{\g \in U_q(su(2)): \;\; \eps(\g) = 0, \; \g^* = S\g \}, \nn\\
\G &=& \{\g \in U_q(su(2)): \;\; \eps(\g) = 1, \; \g^* = S\g \} = e^{\H}
\label{gauge_group}
\err
for  $q \in \R$, and
\berr
\H &=& \{\g \in U_q(su(2)): \;\; \eps(\g) = 0, \; \g^* = S_0\g \}, \nn\\
\G &=& \{\g \in U_q(su(2)): \;\; \eps(\g) = 1, \; \g^* = S_0\g \} = e^{\H}
\label{gauge_group_phase}
\err
for $|q|=1$, where $S_0(u) = q^{H/2} S(u) q^{-H/2}$.
Clearly $\H$ is a subalgebra (without 1) of $U_q(su(2))$, and $\G$
is closed under multiplication. Using the algebra map $j$ 
(\ref{j_map}), $\G$ can be mapped to some real sector of the space of
functions on the fuzzy sphere.

Now consider the following ``gauge'' transformations:
\be
B \rightarrow j(\g_{(1)}) B j(S \g_{(2)}) \qquad \mbox{for} \;\; \g \in \G.
\label{B_gauge}
\ee
It can be checked easily that these transformations preserve the reality
conditions (\ref{B_real}) for both real $q$ and $|q|=1$. 
In terms of components $B = B_a \theta^a r$, this transformation is 
simply (suppressing $j$)
\be
B_a \rightarrow \g_{(1)} B_a S \g_{(2)} = \g \tr B_a,
\ee
which is the rotation of the fields $B_a \in \S_{q,N}$ 
considered as scalar fields\footnote{notice that this is {\em not}
the rotation of the one--form $B$, because 
$\g_{(1)} \xi_i S \g_{(2)} \neq \g\tr \xi_i$}, 
i.e. the rotation $\g \in U_q(su(2))$
does not affect the index $a$ because of (\ref{frame}). In terms of 
the $A_a$ variables, this becomes
\be
A_a \theta^a \rightarrow \g_{(1)} A_a S \g_{(2)}\theta^a + 
\g_{(1)} d( S \g_{(2)})
 = (\g \tr A_a) \theta^a + \g_{(1)} d( S \g_{(2)}),
\label{A_gauge}
\ee
using (\ref{d_0}) and (\ref{frame}).
Hence these transformations are a mixture
of rotations of the components (first term) and 
``pure gauge transformations'' (second term).  Moreover,
the radial and tangential components get mixed. 

To understand these transformations better, consider 
$q=1$. Then we have two transformations of a given gauge field $B_a$, 
the first by conjugation with an unitary element $U \in \S_{q,N}$, and 
the second by (\ref{B_gauge}). We claim that the respective spaces of
inequivalent gauge fields are in fact equivalent. 
Indeed, choose e.g. $a=1$; then there exists a unitary $U \in \S_{q,N}$ 
such that $U^{-1} B_a U$ is a diagonal matrix with real entries.
On the other hand, using a suitable $\g \in \G$ and recalling (\ref{A_decomp}),
one can transform $B_a$ into the form $B_a = \sum_i b_i (x_0)^i$
with real $b_i$, which is again represented
by a diagonal matrix in a suitable basis. Hence at least generically,
the spaces of inequivalent gauge fields are equivalent.

One can also see more intuitively that 
(\ref{A_gauge}) corresponds to an abelian gauge transformation
in the classical limit. 
Consider again $\g(x) = e^{i h(x)}$ with $h(x)^* = -Sh(x)$, 
approximating a smooth function in the limit $N \rightarrow \infty$.
Using properly rescaled variables $x_i$, 
one can see using (\ref{U_A_relation}) that if viewed as an element in 
$U(su(2))$, $\g$ approaches the identity,
that is $\g \tr A_a(x) \rightarrow A_a(x)$ in the classical limit. 
Now write the functions on $\S_{N}$  
in terms of the variables $x_1$ and $x_{-1}$, for example. 
Then (\ref{U_A_relation}) yields
\be
(\one\tens S)\Delta(x_i) = x_i \tens 1 - 1 \tens x_i,
\ee
for $i = \pm 1$, and one can see that  
\be
\g_{(1)} [\l_i, S \g_{(2)}] \approx \partial_i h(x_i)
\ee
in the (flat) classical  limit. Hence (\ref{A_gauge}) indeed 
becomes a gauge transformation in the classical limit. 

To summarize, we found that the set of gauge transformations 
in the noncommutative case is a (real sector of a) quotient of $U_q(su(2))$, 
and can be identified with the space of (real) functions
on $\S_{q,N}$ using the map $j$.
However, the transformation of products of fields is different from the 
classical case. Classically, the gauge group acts on products 
``componentwise'', which means that the coproduct is trivial.
Here, we must assume that $U_q(su(2))$ acts on products of fields
via the $q$--deformed coproduct, so that 
the above actions are invariant under gauge transformations,
by (\ref{invariance_int}). 
In particular, the ``gauge group'' has become a real sector of
a Hopf algebra. Of course, this can be properly implemented on the fields 
only after a ``second quantization'', as in the case of rotation invariance
(see Section \ref{subsec:scalars}). This will be presented in a 
forthcoming paper \cite{ours_2}. This picture is also quite consistent with 
observations of a BRST--like structure in $U_q(so(2,3))$ at roots of unity,
see \cite{thesis}.

Finally, we point out that the above actions are invariant under a global 
$U_q(su(2))$ symmetry, by rotating the frame $\theta^a$.

\sect{Drinfel'd twists and the relation with $D$--branes}


Finally we relate our $q$--deformed fuzzy sphere
to the effective algebra of functions on spherical $D$--branes in the
$SU(2)$  WZW model at level $k$, as determined by Alekseev, 
Recknagel and Schomerus \cite{alekseev}. 
Their result is as follows. The $D$--branes (more precisely their
boundary conditions) are classified by an integer 
$N$ which satisfies\footnote{$N$ is denoted by $2\a$ in \cite{alekseev}}
$0 \leq N \leq k$. The Hilbert space of the associated boundary 
conformal field theory decomposes into irreducible \reps
of the affine Lie algebra $\widehat{su(2)}_k$. One can assign abstract
elements $\{Y^I_i\}_{I,i}$ to the 
boundary vertex operators (primary fields), 
where $I$ ranges from $0$ to $\min(N,k-N)$, 
and $-I \leq i \leq I$. The $\{Y^I_i\}_{i}$ form
irreducible  spin $I$ \reps of the horizontal algebra
$su(2)$, and are interpreted as the analog of spherical harmonics
on the $D$--brane; in particular, there exist only finitely many of them. 
The algebra induced by the OPE of the corresponding
boundary vertex operators is given by \cite{alekseev}
\be
Y^I_i\star Y^J_j = \sum_{K,k} \[\begin{array}{ccc} I&J&K\\ i&j&k \end{array} \]
          \left\{\begin{array}{ccc} I&J&K\\ N/2&N/2&N/2 \end{array} \right\}_q 
             Y^K_k
\qquad \mbox{with} \;\; q=e^{\frac{i\pi }{k+2}},
\label{alekseev_algebra}
\ee    
where the sum goes from $K=0$ to $\min(I+J,k-I-J,N,k-N)$.
This is a finite, noncommutative, quasiassociative algebra $\A$. Here
the first bracket denotes the Clebsch--Gordon coefficients of $SU(2)$,
and the curly brackets denote the $q$--deformed $6J$--symbols of
$U_q(su(2))$. The latter arise from the fusion matrices of the underlying
conformal field theory, which have been known to be related to quantum 
groups for a long time \cite{sierra}. 

In the present paper, we only consider roots of unity $q$ 
which satisfy (\ref{q_phase}). This means that $N \leq k/2$ in the above 
situation, so that
we can only consider a certain subset of the allowed boundary 
conditions here. There will be some qualitative changes in the 
remaining cases, which we do not consider in the present paper.

The reason for the non--associativity of the algebra $\A$ is a
mixing of $q$--deformed and undeformed group theory objects. 
However as was already indicated in \cite{alekseev}, one can sometimes
``twist'' this
algebra using a Drinfeld--twist into an associative one. 
In particular this can be done if $q$ satisfies (\ref{q_phase}), 
in the following way: On the same vector space $\A$, we define a new 
multiplication by
\be
a \tilde\star b := (\F^{-1 (1)}\tr a) \star(\F^{-1 (2)}\tr b) 
            = \star(\F^{-1} \tr(a\tens b)).
\label{star}
\ee 
Here $a,b \in \A$, and 
\be
\F = \F^{(1)}\tens \F^{(2)} \; \in \; U(g)\tens U(g)
\ee
is the Drinfeld twist \cite{Drinfeld_quasi}
in Sweedler--notation. We can ignore 
some fine points here since we only consider certain \reps of $\F$.
The twist relates the
undeformed Clebsch--Gordan coefficients to the deformed ones as follows:
\be
\[\begin{array}{ccc} I&J&K\\ i&j&k' \end{array} \] (g^{(K)})^{k' k}= 
    \[\begin{array}{ccc} I&J&K\\ i'&j'&k' \end{array} \]_q 
     (g_q^{(K)})^{k' k}\;\; \pi^{i'}_i(\F^{(1)}) \pi^{j'}_j(\F^{(2)}) 
\ee
Here we have raised indices using $(g_q^{(K)})^{k' k}$, which 
is the $q$--deformed invariant tensor, and we will
assume that $(g^{(K)})^{k' k} = \d^{k' k}$ (in an orthonormal basis).
It should be noted that even though the abstract element $\F$ exists 
only for generic (more precisely formal) $q$, 
the representations of $\F$ which are needed above 
do exist at roots of unity, assuming the restrictions (\ref{q_phase}) 
on $q$; this is 
because the Clebsch--Gordan decomposition is then still analytic in $q$.
Hence the twisted multiplication rule for the generators $Y^I_i$  is 
\be
Y^I_i \tilde\star Y^J_j 
    = \sum_{K,k} \[\begin{array}{ccc} I&J&K\\ i&j&k' \end{array} \]_q 
      (g_q^{(K)})^{k' k}
    \left\{\begin{array}{ccc} I&J&K\\ N/2&N/2&N/2 \end{array} \right\}_q Y^K_k.
\label{twisted_alekseev}
\ee
As was already pointed out in \cite{alekseev}, this defines an associative
algebra.
We claim that this is precisely the algebra $\S_{q,N}$, which in turn 
is the matrix algebra $Mat(N+1)$. To see this, we 
reconsider the algebra $\S_{q,N}$ from a group--theoretic point of view:

Let now $Y^I_i \in \S_{q,N}$ be an irreducible spin $I$ \rep of $U_q(su(2))$, 
according
to the decomposition (\ref{A_decomp}). In acts on the Fock space $\F_N$ 
(\ref{F_N}), which 
in turn is a spin $N/2$ \rep of $U_q(su(2))$, with a basis of the form
$(A^+ ... \;A^+)_r |0\rangle$. Hence if we denote with 
$\pi(Y^I_i)^r_s$ the matrix which represents the operator $Y^I_i$ 
on $\F_N$,
we can conclude that it is proportional to the Clebsch--Gordan coefficient
of the decomposition $(2I+1) \tens (N+1) \rightarrow (N+1)$;
this is the Wigner--Eckart theorem. Hence in a suitable normalization of 
the basis $Y^I_i$, we can write
\be
\pi(Y^I_i)^r_s =  
 (g_q^{(N/2)})^{r r'} \[\begin{array}{ccc} N/2&I&N/2\\ r'&i&s \end{array} \]_q
 = \[\begin{array}{ccc}I&N/2&N/2\\i&s&r'\end{array} \]_q (g_q^{(N/2)})^{r' r}.
\ee
Therefore the matrix representing the operator 
$Y^I_i Y^J_j$ is given by 
\berr
 \pi(Y^I_i)^r_s \!\!\!\!\!\!\!\!\!\!\!\! && \pi(Y^J_j)^s_t =
  (g_q^{(N/2)})^{r r'} \[\begin{array}{ccc} N/2&I&N/2\\ r'&i&s \end{array} \]_q
  (g_q^{(N/2)})^{s s'} \[\begin{array}{ccc} N/2&J&N/2\\ s'&j&t \end{array} \]_q
    \nn\\
 &=& \sum_K \left\{\begin{array}{ccc} N/2&J&N/2\\I&N/2&K \end{array}\right\}_q
     \[\begin{array}{ccc} I&J&K\\ i&j&k' \end{array} \]_q (g_q^{(K)})^{k' k}
   \[\begin{array}{ccc} K&N/2&N/2\\ k&t&r'\end{array} \]_q (g_q^{(N/2)})^{r' r}
      \nn\\
 &=& \sum_K \left\{\begin{array}{ccc} I&J&K\\ N/2&N/2&N/2\end{array}\right\}_q
    \[\begin{array}{ccc} I&J&K\\ i&j&k' \end{array} \]_q 
                (g_q^{(K)})^{k' k}\; \pi(Y^K_k)^r_t.  \nn\\
\err
Here we used the identity
\be
\left\{\begin{array}{ccc} N/2&J&N/2\\I&N/2&K \end{array}\right\}_q
 = \left\{\begin{array}{ccc} I&J&K\\ N/2&N/2&N/2\end{array}\right\}_q,
\ee
which is proved in \cite{kirill_resh_2}. This calculation 
is represented graphically in Figure 1, which shows that it essentially 
reduces to the definition of the $6j$--symbols.
Therefore the algebra of $\S_{q,N}$ is precisely 
(\ref{twisted_alekseev}), which is a twist of the algebra  
(\ref{alekseev_algebra}) found in \cite{alekseev}. 
In a sense, this twisting is similar to deformation quantization; however, 
$\S_{q,N}$ is a $U_q(su(2))$--module algebra, while (\ref{alekseev_algebra})
is a $U(su(2))$--module algebra. 

\begin{figure}[htpb]
\begin{center}
\epsfxsize=5in
   \hspace{0.1in}
   \epsfbox{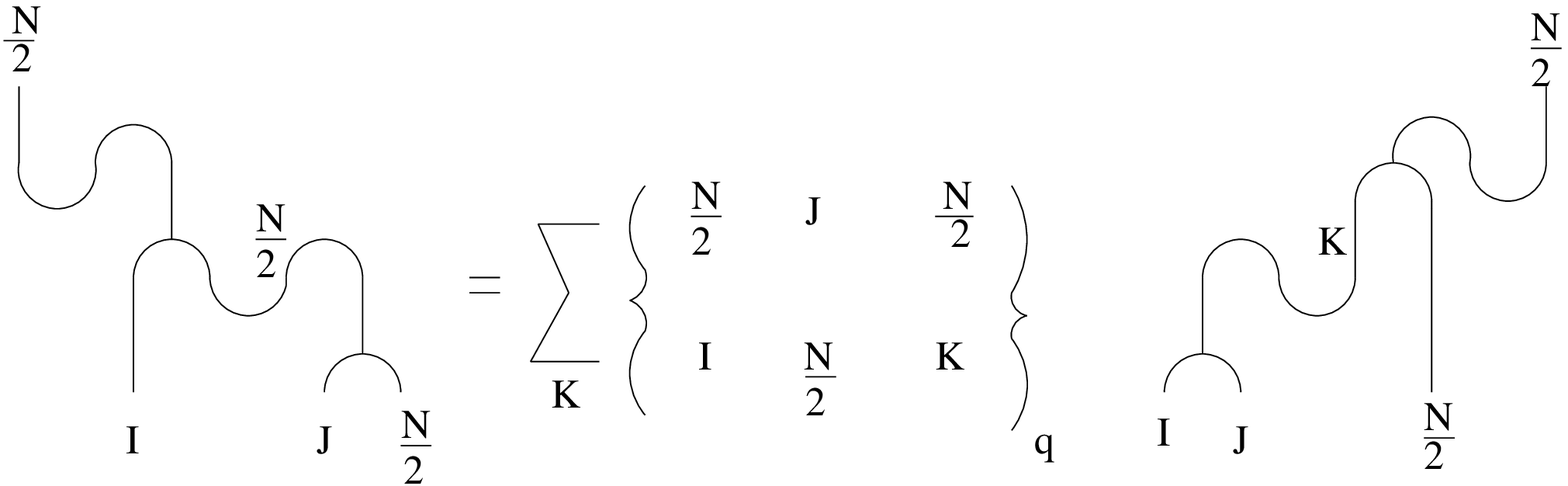}
\end{center}
 \caption{Derivation of the algebra (\ref{twisted_alekseev})}
\label{fig:6jcalc}
\end{figure}

\sect{Summary and outlook}

In this paper, we have studied the $q$--deformed fuzzy sphere $\S_{q,N}$,
which is an associative algebra which is covariant under $U_q(su(2))$, 
for real $q$ and $q$ a phase. In the first 
case, this is the same as the ``discrete series'' of Podle\'s quantum spheres.
We develop the formalism of differential forms and frames, as well as 
integration. We then briefly consider
scalar and gauge field theory on this space. 
It appears that $\S_{q,N}$ is a nice and perhaps the simplest example of
quantum spaces which are covariant under a quantum group.
This makes it particularly well suited for studying field theory,
an endeavour which has proved to be rather difficult on other $q$--deformed
spaces. We are able to write hermitian actions for scalar and gauge fields,
including analogs of Yang--Mills and Chern--Simons actions. 
In particular, the form of the actions for gauge theories suggests a new
type of gauge symmetry, where the role of the gauge group is played by 
$U_q(su(2))$, which can be mapped onto the space of functions on $\S_{q,N}$.
This suggests that formulating field theory on 
quantized spaces which are less trivial than the ones corresponding to a
Moyal product on flat spaces requires new approaches, and may 
lead to interesting new insights. 

The main motivation for doing this is the discovery \cite{alekseev} that 
a quasi--associative twist of $\S_{q,N}$
arises on spherical $D$--branes in the $SU(2)_k$ WZW model, for $q$
a root of unity. In view of this result, 
we hope that the present formalism may be useful 
to formulate a low--energy effective field theory 
induced by open strings ending on the $D$--branes.
This in turn inspires one to consider some kind of second quantization of 
field theories on $\S_{q,N}$, corresponding to a loop expansion and 
many--particle states. It is quite interesting that also from a more 
formal point of view, such a second quantization turns out to be necessary 
for a satisfactory definition of symmetries in such a field theory. 
This will be presented in a future publication \cite{ours_2}.
Moreover, while the question of using either the quasi--associative algebra
(\ref{alekseev_algebra}) or the associative $\S_{q,N}$ may 
ultimately be a matter of taste, the latter does suggest 
certain forms for Lagrangians, induced by the
differential calculus. It would be very interesting 
to compare this with a low--energy effective action 
induced from string theory.

In this paper we have only considered spaces which correspond 
to a subset of the allowed boundary conditions discussed in \cite{alekseev}. 
The remaining cases will show some qualitatively new features, and 
are postponed for future work.

\paragraph{Acknowledgements:}
We would like to thank A. Alekseev, C.-S. Chu, B. Cerchiai, G. Fiore,
J. Pawelczyk, E. Scheidegger and J. Wess for useful and stimulating 
discussions.
H. S. and J. M. are grateful for hospitality at the
Erwin--Schr\"odinger Institut in Vienna, where this work was initiated.
H. G. and J. M. would also like to thank J. Wess and the Max--Planck Institut 
f\"ur Physik in M\"unchen for hospitality and financial support.

\sect{Appendix A: Invariant tensors}

Before giving the explicit forms of the invariant tensors used in this paper,
we briefly explain our conventions and the relation to the 
literature. The quantum spaces in \cite{FRT} and in much of the standard
literature are defined as left $Fun_q(G)$--comodule algebras. This is 
equivalent to right $U_q(g)$--module algebras. However it is more
intuitive to work with left module algebras. This can be achieved using the 
antipode, $u \tr f(x) := f(x) \tl S(u)$; moreover
if $\pi^i_{\;j}(u)$ is the fundamental representation,
then $u \tr x_i = x_j \pi^j_{\;i}(u)$ where $x_i = g_{ij} x^j$. 
However the coproduct then becomes reversed, 
$u \tr f g = (u_2 \tr f) (u_1 \tr g)$. We have incorporated this 
by defining $U_q(g)$ with the reversed coproduct (\ref{coproduct_X})
and antipode. 
This means that our $\RR \in U_q^- \tens U_q^+$ (where $U_q^{\pm}$ denotes
the Borel subalgebras) is obtained from the usual one by 
flipping the tensor components. For example, our 
$\hat R^{+-}_{+-} = \pi^-_{\;\;+}(\RR_1) \pi^+_{\;\;-}(\RR_2) = (q-q^{-1})$ 
in the fundamental \rep of $U_q(su(2))$, where $\pm$ labels the weights. 
Then the characteristic equation and all 
compatibility relations with the invariant tensors have the same form 
as usual, and
are obtained from the standard ones by flipping all horizontal indices.

The $q$--deformed epsilon--symbol (``spinor metric'') for spin 1/2 \reps  
is given by
\be
\eps^{+ -} = q^{-\hf}, \;\; \eps^{- +} = - q^{\hf},
\ee
all other components are zero. The corresponding tensor with lowered indices
is $\eps_{\a \b} = -\eps^{\a \b}$ and satisfies 
$\eps^{\a \b} \eps_{\b \g} = \d^{\a}_{\g}$. In particular, 
$\eps^{\a \b} \eps_{\a \b} = -(q + q^{-1}) = -[2]_q$.

The  $q$--deformed sigma--matrices, i.e. the Clebsch--Gordon coefficients
for $(3) \subset (2) \tens (2)$, are given by 
\berr
\sigma^{+ +}_{1} &=& 1 = \;\sigma^{- -}_{-1}, \nn\\
\sigma^{+ -}_{0} &=& \frac{q^{\hf}}{\sqrt{[2]_q}}, \;\;
\sigma^{- +}_{0} = \frac{q^{-\hf}}{\sqrt{[2]_q}}
\err
in an orthonormal basis, and 
$\sigma^{\a \b}_{i} = \sigma_{\a \b}^{i}$.
They are normalized such that 
$\sum_{\a \b} \sigma_{\a \b}^{i} \sigma^{\a \b}_{j} = \d^i_j$.
That is, they define a unitary map (at least for $q \in \R$).

The $q$--deformed invariant tensor for spin 1 \reps  
is given by
\be
g^{1 -1} = -q^{-1}, \;\; g^{0 0} = 1, \;\; g^{-1 1} = -q,
\ee
all other components are zero.
Then $g_{\a \b} = g^{\a \b}$ satisfies  
$g^{\a \b} g_{\b \g} = \d^{\a}_{\g}$, and 
$g^{\a \b} g_{\a \b} = q^2 +1 + q^{-2} = [3]_q$.

The  Clebsch--Gordon coefficients
for $(3) \subset (3) \tens (3)$, i.e. the $q$--deformed structure 
constants, are given by 
\be
\begin{array}{ll} 
\eps^{1 0}_{1} =  q^{-1}, & \eps^{0 1}_{1} = -q,  \\
\eps^{0 0}_{0} = -(q-q^{-1}), & \eps^{1 -1}_{0} = 1 = -\eps^{-1 1}_{0}, \\
\eps^{0 -1}_{-1} = q^{-1}, & \eps^{-1 0}_{-1} = -q
\end{array}
\label{C_ijk}
\ee
in an orthonormal basis, and $\eps_{i j}^{k} = \eps^{i j}_{k}$.
They are normalized such that 
$\sum_{ij} \eps_{ij}^{n} \eps^{ij}_{m} = [2]_{q^2} \d^n_m$.
Moreover, the following identities hold:
\berr
\eps_{ij}^n g^{jk} &=& \eps^{nk}_i   \label{C_g}  \\
g_{ij} \eps^j_{kl} &=& \eps^j_{ik} g_{jl}  \label{Cg_id} \\
\eps_i^{nk} \eps_k^{lm} - \eps_i^{km} \eps_k^{nl} 
     &=& g^{nl} \d_i^m - \d^n_i g^{lm}   \label{CC_id}   
\err
which can be checked explicitly.
In view of (\ref{Cg_id}), the $q$--deformed totally ($q$--)antisymmetric 
tensor is defined as follws: 
\be
\eps^{ijk} = g^{in} \eps_n^{jk} = \eps^{ij}_n g^{nk}.
\label{q_epsilon}
\ee
It is invariant under the action of $U_q(su(2))$.

\sect{Appendix B: some proofs}

\paragraph{\em Proof of (\ref{dx_i}):}

Using the identity
\be
\one = q^{-2} \hat R + (1+q^{-4})P^- + (1-q^{-6}) P^0,
\label{one_id}
\ee
(\ref{C_g}), (\ref{RRg_braid}),
and the braiding relation (\ref{xxi_braid})
we can calculate the commutation relation of $\Theta$ 
with the generators $x_i$:
\berr
x_i \Theta &=& x_i (x_j \xi_t g^{jt}) \nn\\
  &=& q^{-2} {\hat R}^{kl}_{ij} x_k x_l \xi_t g^{jt} + 
      q^{-2} \L_N \eps^n_{ij} x_n \xi_t g^{jt}
      + \frac{r^2}{[3]} (1-q^{-6}) g_{ij} \xi_t g^{jt} \nn\\
  &=& q^{-2} \Theta x_i + q^{-2} \L_N \eps_{ij}^n x_n \xi_t g^{jt}
              + {r^2}\frac{(1-q^{-6})}{[3]_q} \xi_i \nn\\
  &=&  q^{-2} \Theta x_i + q^{-2} \L_N \eps^{nk}_i x_n \xi_k 
               + {r^2} q^{-3} (q-q^{-1}) \xi_i,  \nn
\err
which yields (\ref{dx_i}).

\paragraph{Proof of (\ref{theta_2}) and (\ref{dtheta}):}

Using (\ref{xixi}), one has
\be
\Theta \xi_i = -q^2 \xi_i \Theta,
\label{theta_xi}
\ee
which implies
$\Theta \Theta = \Theta (x \cdot \xi) = dx \cdot \xi - q^2 \Theta \Theta$, 
hence
$$
(1+q^2) \Theta^2 = dx \cdot \xi.   
$$
On the other hand, (\ref{dx_i}) yields
$$
dx \cdot \xi = - \L_N x_i \eps_j^{kl} \xi_k \xi_l g^{ij}
               - q^3 (q-q^{-1}) \Theta^2,
$$
and combining this it follows that
$$
\Theta^2 = - \frac{q^2 \L_N}{[2]_{q^2}} x_i \xi_k \xi_l \eps^{ikl}.
$$
We wish to relate this to $dx_i dx_j g^{ij}$, which is proportional
to $d \Theta$.
Using  the relations $\eps^{nk}_i x_n \xi_k = -q^{-2} \eps^{nk}_i \xi_n x_k$,
$\Theta = q^{-4} \xi\cdot x$,
(\ref{Cg_id}) and (\ref{CC_id}), one can show that
$$
\eps^{nk}_i x_n \xi_k \eps^{ml}_j x_m \xi_l g^{ij} 
  = \L_N  x_i  \xi_k \xi_l \eps^{ikl} + q^2 \Theta^2
$$
which using (\ref{dx_i}) implies
$$
dx_i dx_j g^{ij} = -  \frac 1{C_N}\;r^2\; \Theta^2.
$$

\paragraph{Proof of $\; d\circ d = 0$ on $\Omega^1_{q,N}$.}

First, we calculate
\berr
[\Theta, d\xi_i] &=& (q^2-1)(q^2+1)\(\xi_i \Theta^2 + 
          \frac{q^{-2} \L_N}{[2]_{q^2}} \eps_i^{kl} \xi_k \xi_l \Theta\) \nn\\
  &=& (q^2-1)(q^{2}+1)  
       \(\xi_i (\ast_H \Theta) - (\ast_H \xi_i) \Theta\) =0  \nn
\err
using (\ref{theta_xi}), (\ref{ast_theta}), and (\ref{star_adj}).
This implies that 
\berr
d(d(f \xi_i)) &=& [\Theta, df \xi_i + f d\xi_i] \nn\\
   &=& [\Theta,  df]_+ \; \xi_i - df [\Theta, \xi_i]_+
       + df d\xi_i + f [\Theta, d\xi_i] \nn\\
  &=& d df + \ast_H( df) \xi_i - 
                   df(d\xi_i + \ast_H(\xi_i)) + df d\xi_i  \nn\\
  &=&  - df \ast_H \xi_i + (\ast_H df) \xi_i = 0 \nn
\err
by (\ref{star_adj}) for any $f \in \S_{q,N}$. This proves 
$d\circ d = 0$ on $\Omega^1_{q,N}$.

\paragraph{Proof of (\ref{star_adj}).}

First, we show that 
\be
(\ast_H \xi_i) \xi_j = \xi_i (\ast_H \xi_j),
\ee
which is equivalent to
$$
\eps^{nk}_i \xi_n \xi_k \xi_j = \xi_i \eps^{nk}_j \xi_n \xi_k.
$$
Now $\Omega^3_{q,N}$ is one--dimensional as module over $\S_{q,N}$,
generated by $\Theta^3$ (\ref{om_3}), which in particular is a singlet
under $U_q(su(2))$. This implies that 
\berr
\eps^{nk}_i \xi_n \xi_k \xi_j &=& 
             (P^0)_{ij}^{rs} \eps^{nk}_r \xi_n \xi_k \xi_s\nn\\
  &=& -\frac{q^6 [2]_{q^2}}{\L_N r^2} \; g_{ij}\;  \Theta^3  \nn\\
  &=& (P^0)_{ij}^{rs} \xi_r \eps^{nk}_s \xi_n \xi_k =\xi_i\eps^{nk}_j \xi_n \xi_k,
\label{cyc_calc}
\err
as claimed. Now (\ref{star_adj}) follows immediately using the
fact that $\ast_H$ is a left--and right $\S_{q,N}$--module map.

\paragraph{Reality structure for $q \in \R$:}
These are the most difficult calculations, and they are needed 
to verify (\ref{obar_xi}) as well. First, we have to show that 
(\ref{xxi_braid}) is compatible with the star structure (\ref{xi_real}).
By a straightforward calculation, one can reduce the problem to
proving (\ref{reality_id}). We verify this by projecting this 
quadratic equation to its spin 0, spin 1, and spin 2 part. 
The first two are easy to check, using (\ref{sxdx_id}) in the spin 1 case. 
To show the spin 2 sector, it is enough to consider (\ref{reality_id})
for $i=j=1$, by covariance. This can be seen e.g. using
$[x_1, \eps_1^{ij} x_i \xi_j] = -q^{-2} \L_N x_1 \xi_1$, which 
in turn can be checked 
using (\ref{one_id}), (\ref{RRC_braid}) and (\ref{C_ijk}). 

Next, we show that (\ref{xixi}) is compatible with the star structure 
(\ref{xi_real}). This can be reduced to 
$$
(q^2 \hat{R} - q^{-2}\hat{R}^{-1})^{k l}_{ij}\; dx_k \; \xi_l
 = q^2(q -q^{-1})\frac{[2]_q C_N}{r^2} 
     (\one + q^2 \hat{R})^{k l}_{ij}) \; dx_k\; dx_l
$$
The spin 0 part is again easy to verify, and the spin 1 part 
vanishes identically (since then $\hat R$ has eigenvalue $-q^{-2}$). 
For the spin 2 part, one can again choose $i=j=1$, and verify it e.g. by 
comparing with the differential of equation (\ref{reality_id}).

\paragraph{Proof of (\ref{cyclic_forms}):}
Since $\Omega^*_{q,N}$ is finitely generated and because of 
(\ref{cyclic_integral}) and $[\Theta^3, f] = 0$, it is enough to consider
$\b = \xi_k$. In this case, the claim reduces to
$$
\xi_i \xi_j \xi_k = S^{-2}(\xi_k) \xi_i \xi_j.
$$
Now $ S^{-2}(\xi_k) = D_k^l \xi_l$, where $D_k^l = \d_k^l q^{2r_l}$
with $r_l = (2,0,-2)$ for $l= (1,0,-1)$, respectively. Since 
$\xi_i \xi_j = \frac 1{[2]_{q^2}}\; \eps_{ij}^n (\eps_{n}^{rs} \xi_r \xi_s)$, 
there remains to show that 
$(\eps_{n}^{rs} \xi_r \xi_s) \xi_k = S^{-2}(\xi_k) (\eps_{n}^{rs} \xi_r \xi_s)$.
By (\ref{cyc_calc}), this is equivalent to
$$
g_{nk} \Theta^3 = D_k^l g_{ln} \Theta^3,
$$
which follows from the definition of $ D_k^l$.

\paragraph{Proof of Lemma \ref{laplace_lemma}:}

Using (\ref{comm_theta2}), (\ref{theta3_f}) and 
$d \Theta^2 = 0$, we have
\berr
d\ast_H d \psi &=& d( \psi \Theta^2 - \Theta^2 \psi)
  = (d\psi)  \Theta^2 - \Theta^2 d\psi \nn\\
  &=& (d\psi)  \Theta^2 + [\Theta^2,\psi] \Theta \nn\\
  &=& (d\psi)  \Theta^2 + (\ast_H d\psi) \Theta.   \label{lapl_calc}
\err
To proceed, we need to evaluate $d\psi_{K}$. 
Because it is an irreducible representation, 
it is enough to consider $\psi_{K} = (x_1)^K$. From 
(\ref{reality_id}) and using 
$\xi_1 x_1 = q^{-2} x_1 \xi_1$, it follows that 
$$
dx_1 x_1 = q^2 x_1 dx_1 - \frac{q^{-2}}{C_N} r^2 x_1 \xi_1,
$$
since $\hat R$ can be replaced by $q^2$ here. By induction, one finds
\be
dx_1 x_1^k = x_1^k \( q^{2k}  dx_1 - 
       [k]_{q^2} \frac{q^{-2}}{C_N}\;r^2 \xi_1 \),
\ee
and by an elementary calculation it follows that
\be
d(x_1^{k+1}) = 
    [k+1]_q x_1^k \(q^k dx_1 - \frac{q^{-2}}{[2]_q C_N} [k]_q r^2 \xi_1 \).
\ee
Moreover, we note that using (\ref{star_adj}) 
\be
(\xi_i \Theta + \ast_H \xi_i) \Theta =
  \xi_i (\ast_H\Theta) + (\ast_H\xi_i) \Theta = 2 (\ast_H\xi_i) \Theta 
  = -\frac 2{r^2} x_i \Theta^3.
\ee
The last equality follows easily from (\ref{cyc_calc}) and (\ref{theta_xi}).
Similarly
\be
(dx_i \Theta + \ast_H dx_i) \Theta = 2 \ast_H d x_i \Theta = 2 dx_i \Theta^2.
\ee
Now we can continue (\ref{lapl_calc}) as  
\berr
d\ast_H d x_1^{K} &=& (dx_1^{K-1} \;\Theta + \ast_H dx_1^{K-1}) \Theta \nn\\
  &=& [K]_q  x_1^{K-1} \(2 q^{K-1} dx_1 \Theta^2 -    
     2 \frac{q^{-2}}{[2]_q C_N} [K-1]_q x_1 \Theta^3\).
\err
Finally it is easy to check that
\be
dx_i \Theta^2 = -\frac 1{C_N} x_i \Theta^3,
\ee
and after a short calculation one finds (\ref{scalar_laplace}).

\end{document}